\documentclass[prb,twocolumn,showpacs,preprintnumbers,amsmath,amssymb,superscriptaddress]{revtex4-2}

\usepackage{graphicx}
\usepackage{dcolumn}
\usepackage{bm}
\usepackage{amsthm}
\usepackage{amsmath}
\usepackage{amssymb}

\begin{document}

\title{Neutron study of magnetic correlations in rare-earth-free Mn--Bi magnets}

\author{Artem Malyeyev}\email{artem.malyeyev@uni.lu}
\affiliation{Department of Physics and Materials Science, University of Luxembourg, 162A~Avenue de la Fa\"iencerie, L-1511 Luxembourg, Grand Duchy of Luxembourg}

\author{Ivan Titov}
\affiliation{Department of Physics and Materials Science, University of Luxembourg, 162A~Avenue de la Fa\"iencerie, L-1511 Luxembourg, Grand Duchy of Luxembourg}

\author{Philipp Bender}
\altaffiliation{Now at:~Heinz Maier-Leibnitz Zentrum (MLZ), Technische Universit\"at M\"unchen, D-85748 Garching, Germany}
\affiliation{Department of Physics and Materials Science, University of Luxembourg, 162A~Avenue de la Fa\"iencerie, L-1511 Luxembourg, Grand Duchy of Luxembourg}

\author{Mathias Bersweiler}
\affiliation{Department of Physics and Materials Science, University of Luxembourg, 162A~Avenue de la Fa\"iencerie, L-1511 Luxembourg, Grand Duchy of Luxembourg}

\author{Vitaliy Pipich}
\affiliation{Forschungszentrum J\"ulich GmbH, J\"ulich Centre for Neutron Science (JCNS) at Heinz Maier-Leibnitz Zentrum (MLZ), Lichtenbergstra{\ss}e~1, D-85748 Garching, Germany}

\author{Sebastian M\"uhlbauer}
\affiliation{Heinz Maier-Leibnitz Zentrum (MLZ), Technische Universit\"at M\"unchen, D-85748 Garching, Germany}

\author{Semih Ener}
\affiliation{Institute of Materials Science, Technical University of Darmstadt, D-64287 Darmstadt, Germany}

\author{Oliver Gutfleisch}
\affiliation{Institute of Materials Science, Technical University of Darmstadt, D-64287 Darmstadt, Germany}

\author{Andreas Michels}\email{andreas.michels@uni.lu}
\affiliation{Department of Physics and Materials Science, University of Luxembourg, 162A~Avenue de la Fa\"iencerie, L-1511 Luxembourg, Grand Duchy of Luxembourg}

\begin{abstract}
We report the results of an unpolarized small-angle neutron scattering (SANS) study on Mn--Bi-based rare-earth-free permanent magnets. The magnetic SANS cross section is dominated by long-wavelength transversal magnetization fluctuations and has been analyzed in terms of the Guinier-Porod model and the distance distribution function. This provides the radius of gyration which, in the remanent state, ranges between about $220$$-$$240 \, \mathrm{nm}$ for the three different alloy compositions investigated. Moreover, computation of the distance distribution function in conjunction with results for the so-called $s$-parameter obtained from the Guinier-Porod model indicate that the magnetic scattering of a Mn$_{45}$Bi$_{55}$ sample has its origin in slightly shape-anisotropic structures.
\end{abstract}

\maketitle

\section{Introduction}
\label{introduction}

Permanent magnets are the subject of an intense worldwide research effort, which is due to their technological relevance as integral components in many electronics devices or motors~\cite{Gutfleisch2011,Riba2016}. Currently, the worldwide permanent magnet market is dominated by two classes of magnets:~(i)~High-performance Nd-Fe-B with a maximum energy product of $(BH)_{\mathrm{max}} \cong 400 \, \mathrm{kJm}^{-3}$ at $300 \, \mathrm{K}$, and (ii)~low-performance ferrite magnets with a $(BH)_{\mathrm{max}} \lesssim 40 \, \mathrm{kJm}^{-3}$. There is a need for a medium-performance and cost-effective material working at temperatures as high as $500 \, \mathrm{K}$ (typical operating temperature of motors), where the $(BH)_{\mathrm{max}}$ of ternary Nd-Fe-B is unacceptably low; in other words, a low-cost permanent magnet material is required which may replace Nd-Fe-B in such applications where the full potential of the latter is not exploited. Rare-earth-free Mn-based permanent magnets are a prime candidate for filling the gap between Nd-Fe-B and the ferrites~\cite{Coey2012}. Mn-based magnets in general~\cite{Coey2014,Ener2015,Jian2015,gutfleisch2020} and the low-temperature phase of Mn--Bi binary alloy in particular~\cite{Park2014,Ly2014,Baker2015,Chen2015,Kim2017,Nguyen2018} have received a lot of attention lately, not the least because of a positive temperature coefficient of the magnetic anisotropy rendering high-temperature applications attractive~\cite{Chen2015}.

Most of the published studies on Mn--Bi-based magnets have focused on integral measurement techniques and on engineering aspects~\cite{Nguyen2014,Cui2014b,Poudyal2016,Mitsui2016,Xiang2018,Xiang2018b,Janotov2018,Cao2019}. Yet, the macroscopic magnetic properties arise, at least partly, from spatial variations in the magnitude and orientation of the magnetization vector field $\mathbf{M}(\mathbf{r})$ on a mesoscopic length scale (a few nm up to the micron scale). Therefore, a deeper understanding of the correlations and long-wavelength magnetization fluctuations is of paramount importance both from the basic science point of view as well as from a materials science perspective aiming to optimize the properties of the material.

In this paper we report the results of unpolarized magnetic small-angle neutron scattering (SANS) experiments on cold-compacted isotropic Mn--Bi magnets. The magnetic SANS method is ideally suited to characterize the magnetic structure and interactions on the mesoscopic length scale, since it provides information on both variations of the magnitude and orientation of the magnetization $\mathbf{M}(\mathbf{r})$ in the bulk of the material. This technique has previously been applied to study e.g.\ the structures of magnetic nanoparticles~\cite{disch2012,guenther2014,bender2015,bender2018jpcc,bender2018prb,oberdick2018,krycka2019,benderapl2019,bersweiler2019,zakutna2020,laura2020}, soft magnetic nanocomposites and complex alloys~\cite{suzuki2007,herr08pss,mettusprm2017,mirebeau2018,schroeder2020,bersweiler2020,oba2020}, proton domains~\cite{michels06a,aswal08nim,noda2016}, magnetic steels~\cite{bischof07,bergner2013,Pareja:15,gilbert2016,shu2018}, or Heusler-type alloys~\cite{bhatti2012,runov2006,michelsheusler2019,leighton2019,sarkar2020} (see Ref.~\onlinecite{Muhlbauer2019} for a recent review on magnetic SANS). When conventional SANS is extended by very small-angle neutron scattering, as in the present case, then the accessible real-space length scale may range from a few nanometers up to the micron regime. Here, we aim to estimate the characteristic size of microstructural-defect-induced spin perturbations in the polycrystalline microstructure of Mn--Bi magnets.

The paper is organized as follows:~Section~\ref{experiments} furnishes the details on the sample synthesis and on the neutron experiments. Section~\ref{sans} displays the expressions for the SANS cross section, the generalized Guinier-Porod model, and the distance distribution function. Section~\ref{results} then presents and discusses the neutron results. Finally, Sec.~\ref{conclusion} summarizes the main findings of this study. We refer to the Supplemental Material for additional supporting information~\cite{suppmaterial}.

\section{Experimental}
\label{experiments}

All Mn--Bi samples were synthesized by using conventional melting and milling, similar to Refs.~\onlinecite{Chen2016,Muralidhar2017,Chen2015}. Initial ingots were prepared by arc melting high-purity elements ($99.8 \, \%$ for Mn and $99.99 \, \%$ for Bi) and were annealed under Ar atmosphere for $24 \, \mathrm{h}$ at $300 \, ^\circ$C followed by quenching in water at room temperature. Subsequently, the resulting ingots were hand crushed under N$_2$ atmosphere into powder (with a particle size $< 60 \, \mu$m) and ball milled for $2 \, \mathrm{h}$ in hexane with a ball-to-powder weight ratio of $1$$:$$10$ at $150 \, \mathrm{RPM}$~\cite{Cao2019}. The ball milled powder was washed in ethanol, magnetically separated, dried, and cold compacted at a pressure of $\sim 1.0 \, \mathrm{GPa}$ into $10 \times 5 \times 1 \, \mathrm{mm}$ pellets. Magnetization isotherms were recorded using a vibrating sample magnetometer (Cryogenic, $\mu_0 H_{\mathrm{max}} = 14 \, \mathrm{T}$). For more details on sample preparation and characterization using x-ray diffraction and scanning electron microscopy see Ref.~\onlinecite{Chen2016}.

\begin{figure}[tb!]
\resizebox{1.0\columnwidth}{!}{\includegraphics{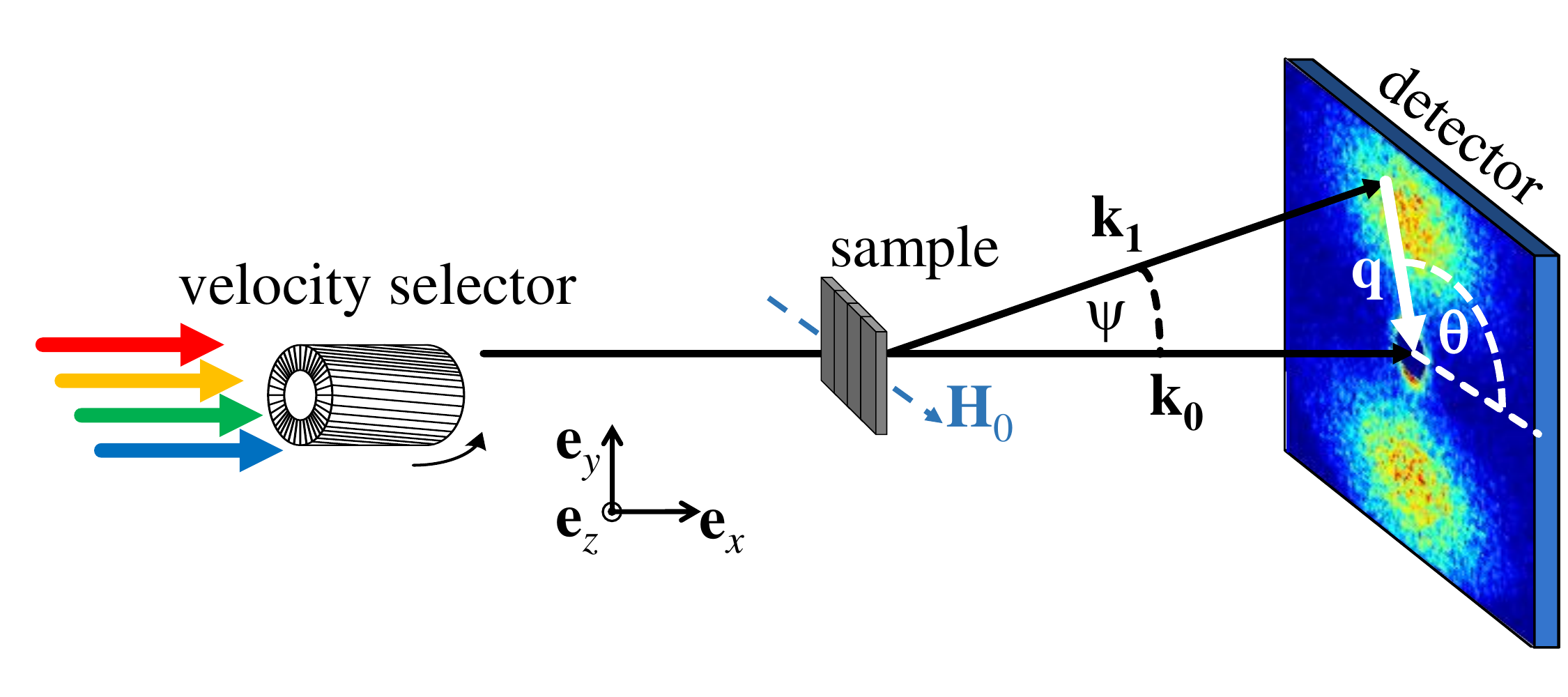}}
\caption{Sketch of the SANS setup. The external magnetic field to the sample, $\mathbf{H}_0$, is applied perpendicular to the incident neutron beam. The scattering vector $\mathbf{q}$ is defined as the difference between the wave vectors of the incident ($\mathbf{k}_0$) and the scattered ($\mathbf{k}_1$) neutrons, i.e., $\mathbf{q} = \mathbf{k}_0 - \mathbf{k}_1$. Its magnitude for elastic scattering, $q = (4 \pi / \lambda) \sin(\psi/2)$, depends on the mean wavelength $\lambda$ of the neutrons (selected by the velocity selector) and on the scattering angle $\psi$. The angle $\theta$ specifies the orientation of $\mathbf{q}$ on the two-dimensional position-sensitive detector. In the small-angle approximation the component of $\mathbf{q}$ along $\mathbf{k}_0$ is neglected, so that $\mathbf{q} \cong \{0, q_y, q_z \} = q \{ 0, \sin\theta, \cos\theta \}$ for $\mathbf{H}_0 \perp \mathbf{k}_0$.}
\label{sans_setup}
\end{figure}

The unpolarized SANS experiments were conducted at room temperature at the very small-angle neutron scattering instrument KWS-3~\cite{Pipich2015} at the Heinz Maier-Leibnitz Zentrum (MLZ), Garching, Germany. We employed samples grinded down to a thickness of $0.1 \, \mathrm{mm}$. The external magnetic field $\mathbf{H}_0$ was applied perpendicular to the incident neutron beam ($\mathbf{H}_0 \perp \mathbf{k}_0$), and a mean wavelength of $\lambda = 12.8 \, \text{\AA}$ with a bandwidth of $\Delta\lambda / \lambda \cong 10 \, \%$ (FWHM) was chosen; see Fig.~\ref{sans_setup} for a sketch of the neutron setup. The covered momentum transfer ranges between about $0.002 \, \mathrm{nm}^{-1} \lesssim q \lesssim 0.2 \, \mathrm{nm}^{-1}$. The neutron experiments were performed by first applying a field of $2.2 \, \mathrm{T}$ and then reducing the field following the magnetization curve (compare Fig.~\ref{VSMXRD_both}). SANS data reduction (correction for background scattering, transmission, detector efficiency) was carried out using the QTI-SAS software package~\cite{qtisas}. Additional neutron measurements on the instrument SANS-1 at MLZ~\cite{MUHLBAUER2016297} were performed (see Ref.~\onlinecite{suppmaterial} for details).

\section{SANS cross section, generalized Guinier-Porod model, and distance distribution function}
\label{sans}

In this section the expressions for the unpolarized SANS cross section, for the Guinier-Porod model, as well as for the distance distribution function are displayed. For more background details on magnetic SANS the reader is referred to Refs.~\onlinecite{michels2014review,Muhlbauer2019}.

\subsection{Unpolarized SANS cross section}

For the perpendicular scattering geometry ($\mathbf{H}_0 \perp \mathbf{k}_0$) the elastic unpolarized SANS cross section $d \Sigma / d \Omega$ at momentum-transfer vector $\mathbf{q}$ can be written as~\cite{michels2014review,Muhlbauer2019}:

\begin{equation} \label{sigmaPerpUnpol}
\begin{split}
\frac{d \Sigma}{d \Omega}(\mathbf{q}) = & \frac{8 \pi^3}{V} b_H^2 \left( b_H^{-2}\right. |\widetilde{N}|^2   
 +|\widetilde{M}_x|^2 + |\widetilde{M}_y|^2 \cos^2\theta +\\
 &+|\widetilde{M}_z|^2 \sin^2\theta  -\Big. (\widetilde{M}_y \widetilde{M}_z^{\ast} + \widetilde{M}_y^{\ast} \widetilde{M}_z) \sin\theta \cos\theta \Big) ,
\end{split}
\end{equation}
where $V$ is the scattering volume, $b_H = 2.91 \times 10^{8} \, \mathrm{A^{-1} m^{-1}}$ is the magnetic scattering length in the small-angle regime, $\widetilde{N}(\mathbf{q})$ and $\widetilde{\mathbf{M}}(\mathbf{q})$ denote, respectively, the Fourier transforms of the nuclear scattering length density $N(\mathbf{r})$ and of the magnetization vector field $\mathbf{M}(\mathbf{r})$, the angle $\theta$ is measured between $\mathbf{q}$ and $\mathbf{H}_0$, and the asterisk ``$\,^{\ast}\,$'' marks the complex-conjugated quantity. We would like to emphasize that the magnetization of a bulk ferromagnet is a function of the position $\mathbf{r} = \{ x, y, z \}$ inside the material, i.e., $\mathbf{M} = \mathbf{M}(x, y, z)$, and that, consequently, $\widetilde{\mathbf{M}} = \widetilde{\mathbf{M}}(q_x, q_y, q_z)$. However, the above Fourier components represent projections into the plane of the two-dimensional detector, i.e., the $q_y$-$q_z$-plane for $\mathbf{H}_0 \perp \mathbf{k}_0$ ($q_x \cong 0$) (compare Fig.~\ref{sans_setup}). This shows that SANS predominantly measures correlations in the plane perpendicular to the incident neutron beam.

In our data analysis we subtract the total nuclear and magnetic SANS cross section at the highest available field from the data at lower fields. This eliminates the nuclear SANS contribution in Eq.~(\ref{sigmaPerpUnpol}) and yields the purely magnetic SANS cross section $d \Sigma_M / d \Omega$; to be more precise, the subtraction procedure results in a magnetic SANS cross section which depends on the \textit{differences} of the magnetization Fourier components at the two fields considered, e.g., $\Delta |\widetilde{M}_x|^2 = |\widetilde{M}_x|^2(H_0) - |\widetilde{M}_x|^2(H_{\mathrm{max}})$ (and similarly for the other Fourier components). The field dependence of the transversal magnetization Fourier components $\widetilde{M}_x$ and $\widetilde{M}_y$ is different from, and usually much larger than, the longitudinal component $\widetilde{M}_z$ (see Fig.~8 in \cite{michels2014jmmm}); more specifically, $\widetilde{M}_{x,y}$ are usually larger at lower field than at higher field, whereas $\widetilde{M}_z$ may weakly increase with increasing field. Effectively, for Mn--Bi, this entails that the difference SANS cross section is non-negative at all $\mathbf{q}$ and $H_0$ investigated.

\subsection{Generalized Guinier-Porod model}

The magnetic SANS cross section $d \Sigma_M / d \Omega$ was analyzed in terms of the generalized Guinier-Porod model, developed by Hammouda~\cite{hammouda2010a} in order to describe the $2 \pi$-azimuthally-averaged scattering from both spherical and nonspherical objects. The model is purely empirical and essentially decomposes the $I(q) = \frac{d \Sigma_M}{d \Omega}(q)$ curve into a Guinier region for $q \leq q_1$ and into a Porod region for $q \geq q_1$. Both parts of the scattering curve are then joined by demanding the continuity of the Guinier and Porod laws (and of their derivatives) at $q_1$; more specifically~\cite{hammouda2010a}:
\begin{eqnarray}
\label{gpmodeleq1}
I(q) &=& \frac{G}{q^s} \exp\left( - \frac{q^2 R_G^2}{3-s} \right) \hspace{0.25cm} \mathrm{for} \hspace{0.25cm} q \leq q_1 , \\
I(q) &=& \frac{D}{q^n} \hspace{0.25cm} \mathrm{for} \hspace{0.25cm} q \geq q_1 ,
\end{eqnarray}
where the scaling factors $G$ and $D$, the Guinier radius $R_G$, the dimensionality factor $s$, and the Porod power-law exponent $n$ are taken as independent parameters. From the continuity of the Guinier and Porod functions and their derivatives it follows that:
\begin{eqnarray}
\label{gpmodeleq2a}
q_1 &=& \frac{1}{R_G} \left[ \frac{(n-s)(3-s)}{2} \right]^{1/2} , \\
D &=& G q_1^{n-s} \exp\left( - \frac{q_1^2 R_G^2}{3-s} \right) ,
\label{gpmodeleq2}
\end{eqnarray} 
where $n>s$ and $s<3$ must be satisfied. Note that $q_1$ is not a fitting parameter, but an internally computed value [via Eq.~(\ref{gpmodeleq2a})]. For a dilute set of homogeneous {\it spherical} particles with sharp interfaces one expects $s=0$, $n=4$, and $R_G^2 = \frac{3}{5} R^2$, where $R$ is the particle radius.

\subsection{Distance distribution function}

In addition to the above analysis using the generalized Guinier-Porod model, we have model-independently calculated the distance distribution function~\cite{bender2017}:
\begin{equation}
\label{pvonrfunc}
p(r) = r^2 \int_0^{\infty} \frac{d \Sigma_M}{d \Omega}(q) j_0(q r) q^2 dq ,
\end{equation}
where $j_0(qr) = \sin(qr)/(qr)$ denotes the zeroth-order spherical Bessel function. This provides information on the characteristics (e.g., size and shape) of the scattering objects~\cite{svergun03,glatter2006}, and on the presence of interparticle correlations~\cite{glatter1996,glatter2011}.

\section{Results and Discussion}
\label{results}

\begin{figure}[tb!]
\includegraphics[width=1.0\linewidth]{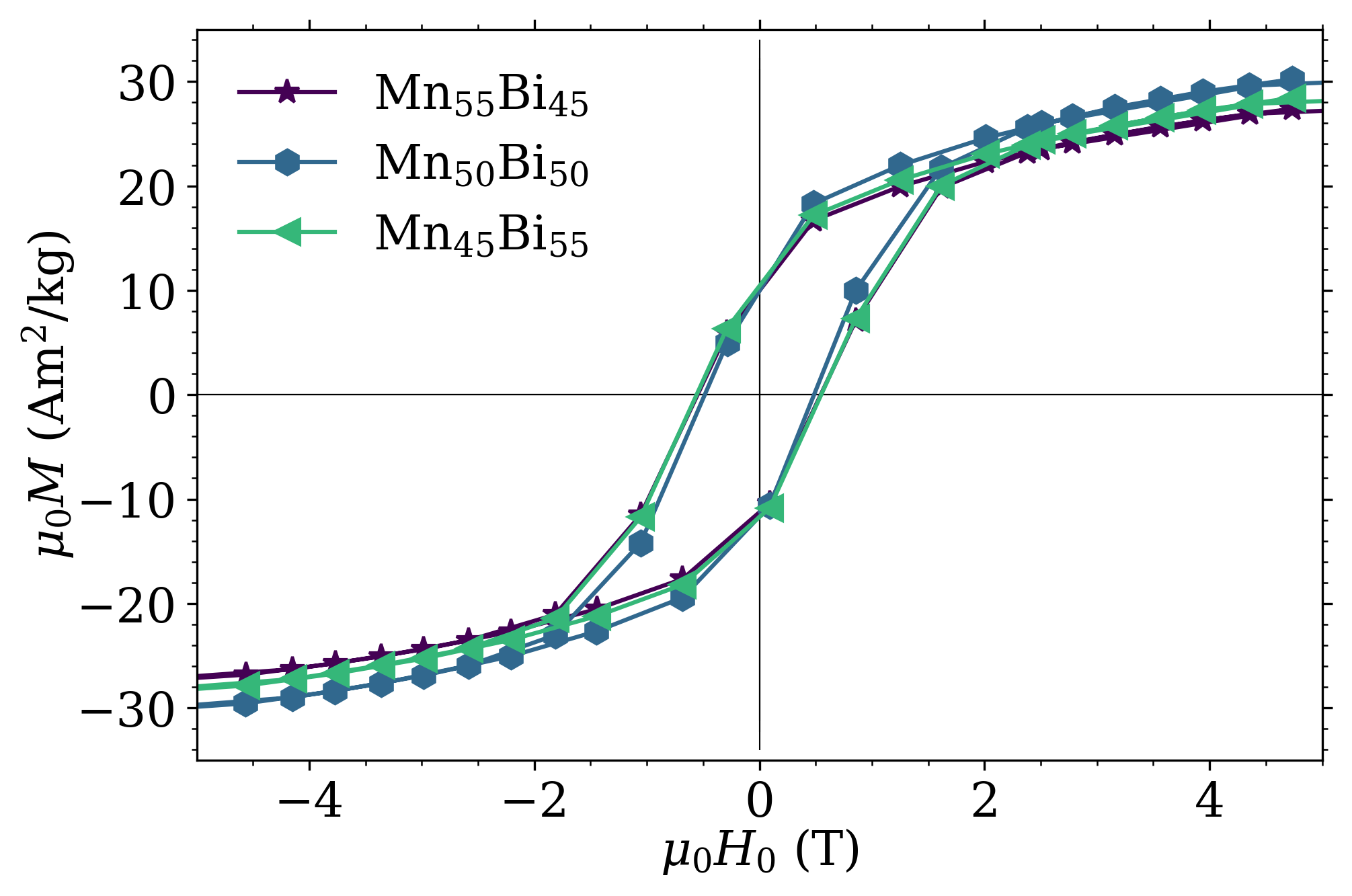}
\caption{Room-temperature magnetization curves $M(H_0)$ of various Mn--Bi samples (see inset).}
\label{VSMXRD_both}
\end{figure}

The room-temperature magnetization curves of the Mn--Bi samples are shown in Fig.~\ref{VSMXRD_both}. The coercivity $H_c$ of the samples is found to be between $0.47$$-$$0.56 \, \mathrm{T}$ for all compositions, while the saturation magnetization $M_s$ varies from about $33 \, \mathrm{Am^2kg^{-1}}$ (Mn$_{55}$Bi$_{45}$) to $36 \, \mathrm{Am^2kg^{-1}}$ (Mn$_{50}$Bi$_{50}$) to $34 \, \mathrm{Am^2kg^{-1}}$ (Mn$_{45}$Bi$_{55}$). These values fall short of the theoretical saturation magnetization of the low-temperature Mn--Bi phase ($80 \, \mathrm{Am^2kg^{-1}}$) and indicate a magnetic content of $\sim 40$$-$$45 \, \%$. A field larger than $1.8 \, \mathrm{T}$ is sufficient to close the hysteresis loop and to reach the reversible part of the $M(H_0)$~curve. This is important because in the neutron-data analysis the measurement at $2.2 \, \mathrm{T}$ is used for subtraction to eliminate the nuclear scattering.

\begin{figure*}[tb!]
\vspace*{-10pt}
\centering
\includegraphics[width=1.0\linewidth]{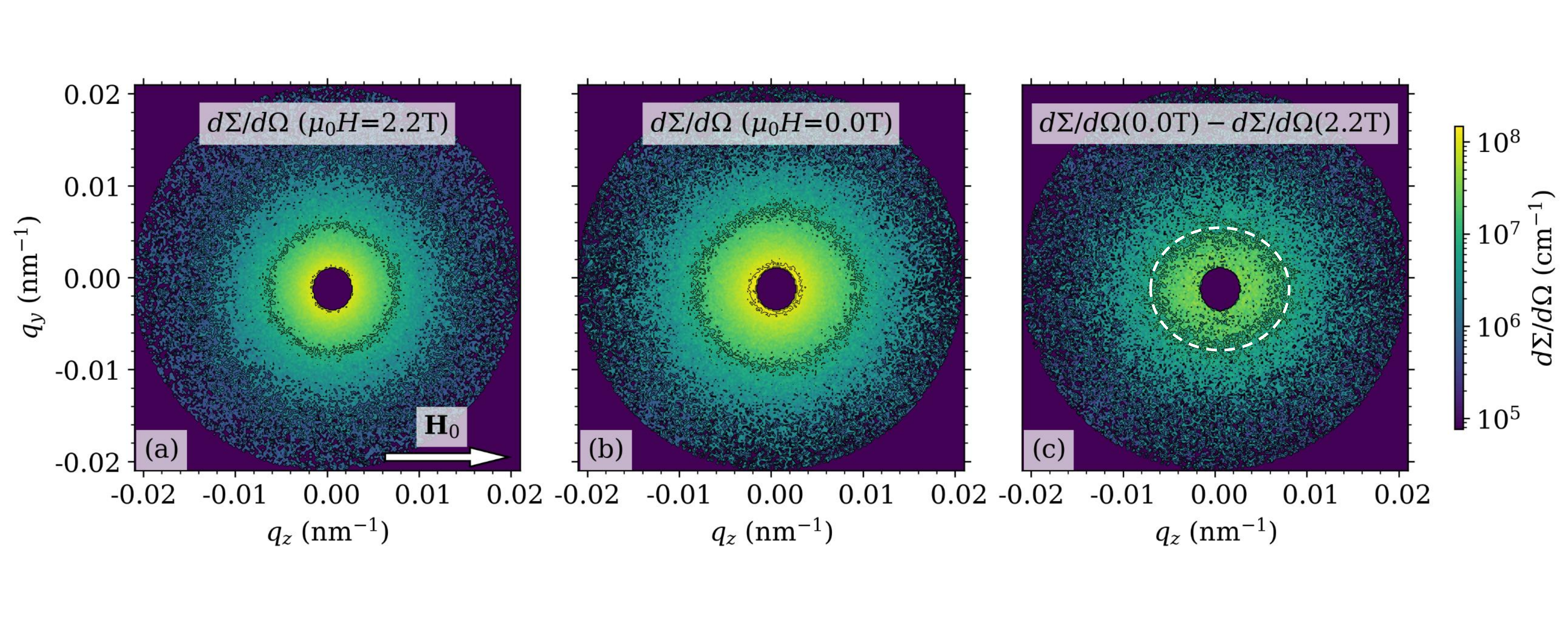}
\caption{Illustration of the neutron data analysis procedure. Shown is the total two-dimensional SANS cross section of a Mn$_{55}$Bi$_{45}$ rare-earth-free permanent magnet ($\mathbf{H}_0 \perp \mathbf{k}_0$; logarithmic color scale). (a)~Total (nuclear and magnetic) SANS cross section $d \Sigma / d \Omega$ at $\mu_0 H_0 = 2.2 \, \mathrm{T}$ ($\mathbf{H}_0$ is horizontal in the plane, see inset). (b)~$d \Sigma / d \Omega$ at remanence ($0 \, \mathrm{T}$). (c)~Magnetic (difference) SANS cross section $d \Sigma_M / d \Omega$ at remanence. The dashed white line emphasizes the slight elongation of $d \Sigma_M / d \Omega$ along $\mathbf{H}_0$~\cite{suppmaterial}.}
\label{2D_method}
\end{figure*}

\begin{figure*}[tb!]
\centering
\includegraphics[width=1.0\linewidth]{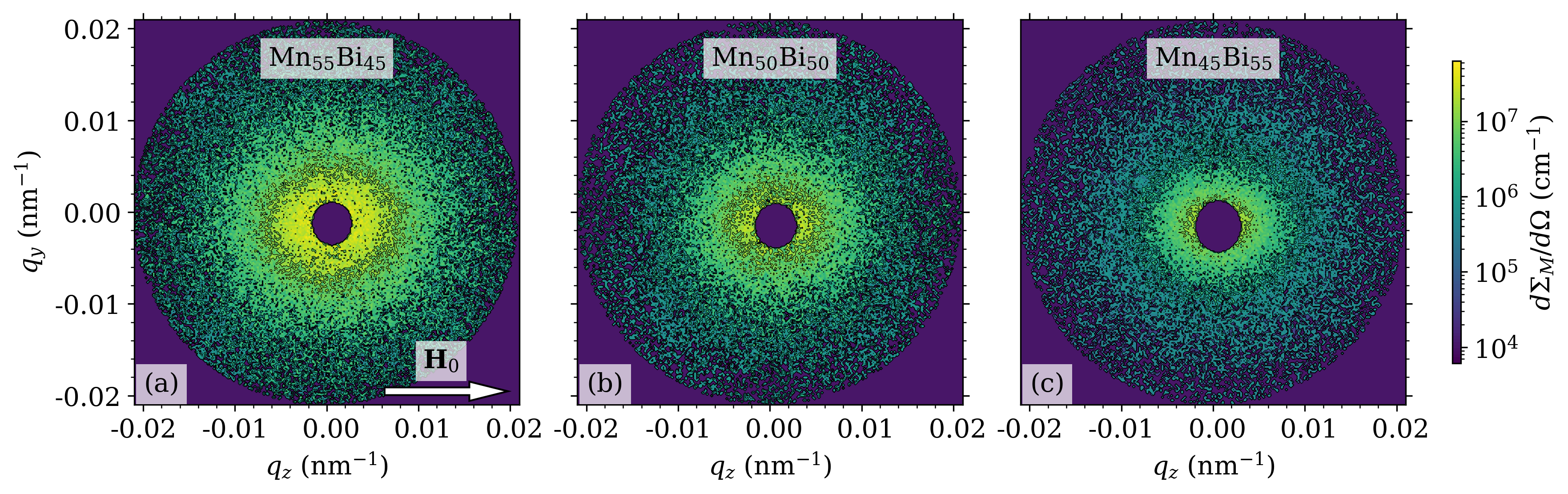}
\caption{Two-dimensional magnetic (difference) SANS cross section $d \Sigma_M / d \Omega$ of Mn--Bi rare-earth-free permanent magnets at the remanent state ($\mathbf{H}_0 \perp \mathbf{k}_0$; logarithmic color scale).}
\label{2D_MnBix3}
\end{figure*}

Figure~\ref{2D_method} illustrates part of our neutron data analysis procedure, which is based on the subtraction of the total $d \Sigma / d \Omega$ at the highest field [Fig.~\ref{2D_method}(a)] from data at lower fields [Fig.~\ref{2D_method}(b)]. This eliminates the strong and presumably isotropic nuclear SANS contribution [compare Eq.~(\ref{sigmaPerpUnpol})] and provides access to the purely magnetic SANS cross section $d \Sigma_M / d \Omega$ [Fig.~\ref{2D_method}(c)]~\cite{michels2014review, Muhlbauer2019}. As can be seen in Fig.~\ref{2D_MnBix3}, the in this way obtained $d \Sigma_M / d \Omega$ are anisotropic, elongated along the direction {\it parallel} to the applied magnetic field $\mathbf{H}_0$; compare the sector-averaged data in~\cite{suppmaterial}. By comparison to the expression for $d \Sigma / d \Omega$ in the $\mathbf{H}_0 \perp \mathbf{k}_0$~geometry [Eq.~(\ref{sigmaPerpUnpol})] this angular anisotropy can be related to the {\it transversal} Fourier component $|\widetilde{M}_y|^2 \cos^2\theta$ in $d \Sigma_M / d \Omega$. The feature is observable for all Mn--Bi samples in the remanent state (Fig.~\ref{2D_MnBix3}), and it suggests the presence of long-range spin-misalignment correlations on a real-space length scale of at least a few ten to a few hundreds of nanometers.

\begin{figure}[tb!]
\centering
\includegraphics[width=1.0\linewidth]{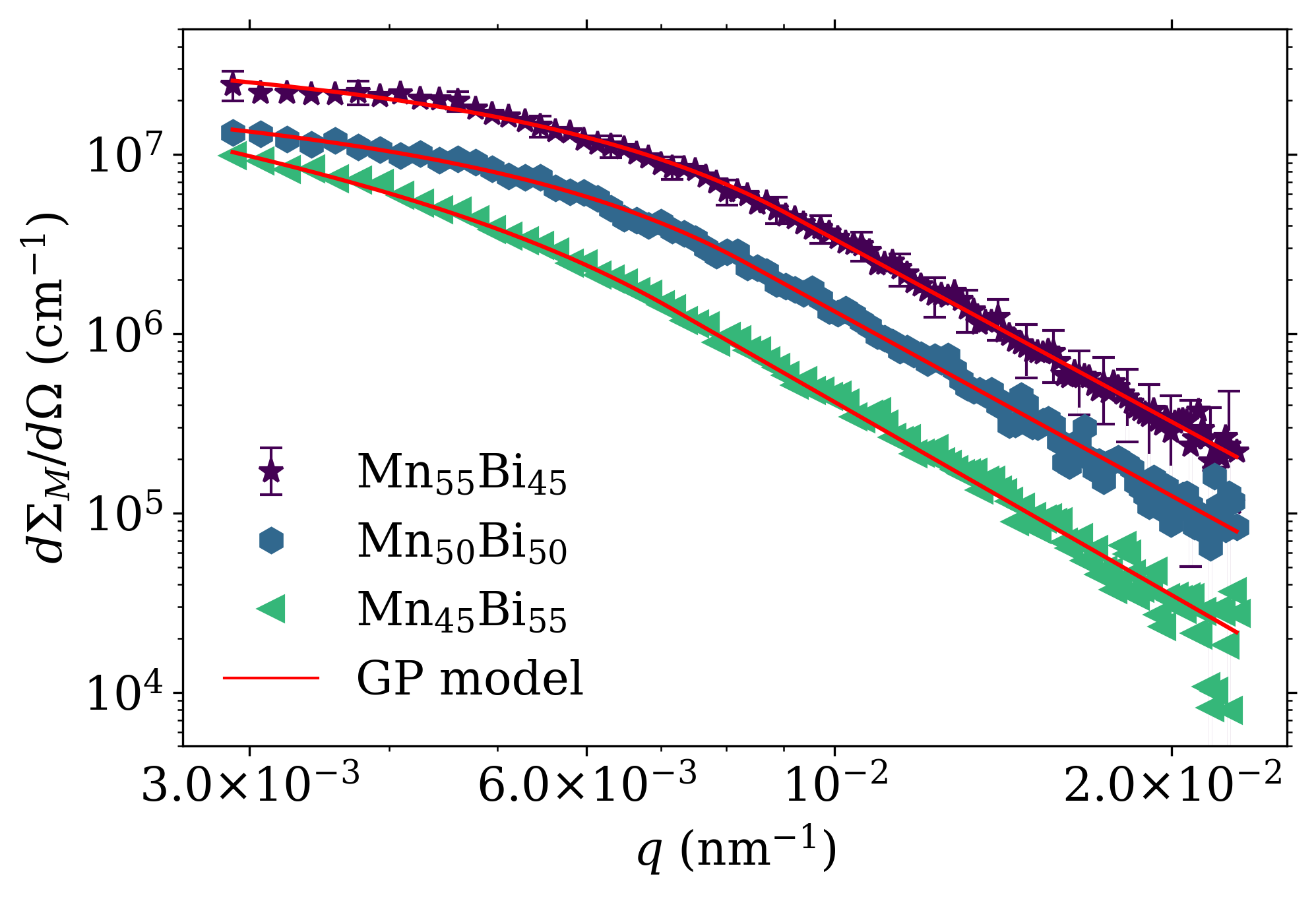}
\caption{$2\pi$-azimuthally-averaged $d \Sigma_M / d \Omega$ of Mn--Bi rare-earth-free permanent magnets in the remanent state ($\mathbf{H}_0 \perp \mathbf{k}_0$; log-log scale). Solid lines:~Fit to the generalized Guinier-Porod model [Eqs.~(\ref{gpmodeleq1})$-$(\ref{gpmodeleq2})]. Error bars are selectively shown only for the Mn$_{55}$Bi$_{45}$ sample.}
\label{1dguinierfit}
\end{figure}

\begin{figure}[tb!]
\centering
\includegraphics[width=1.0\linewidth]{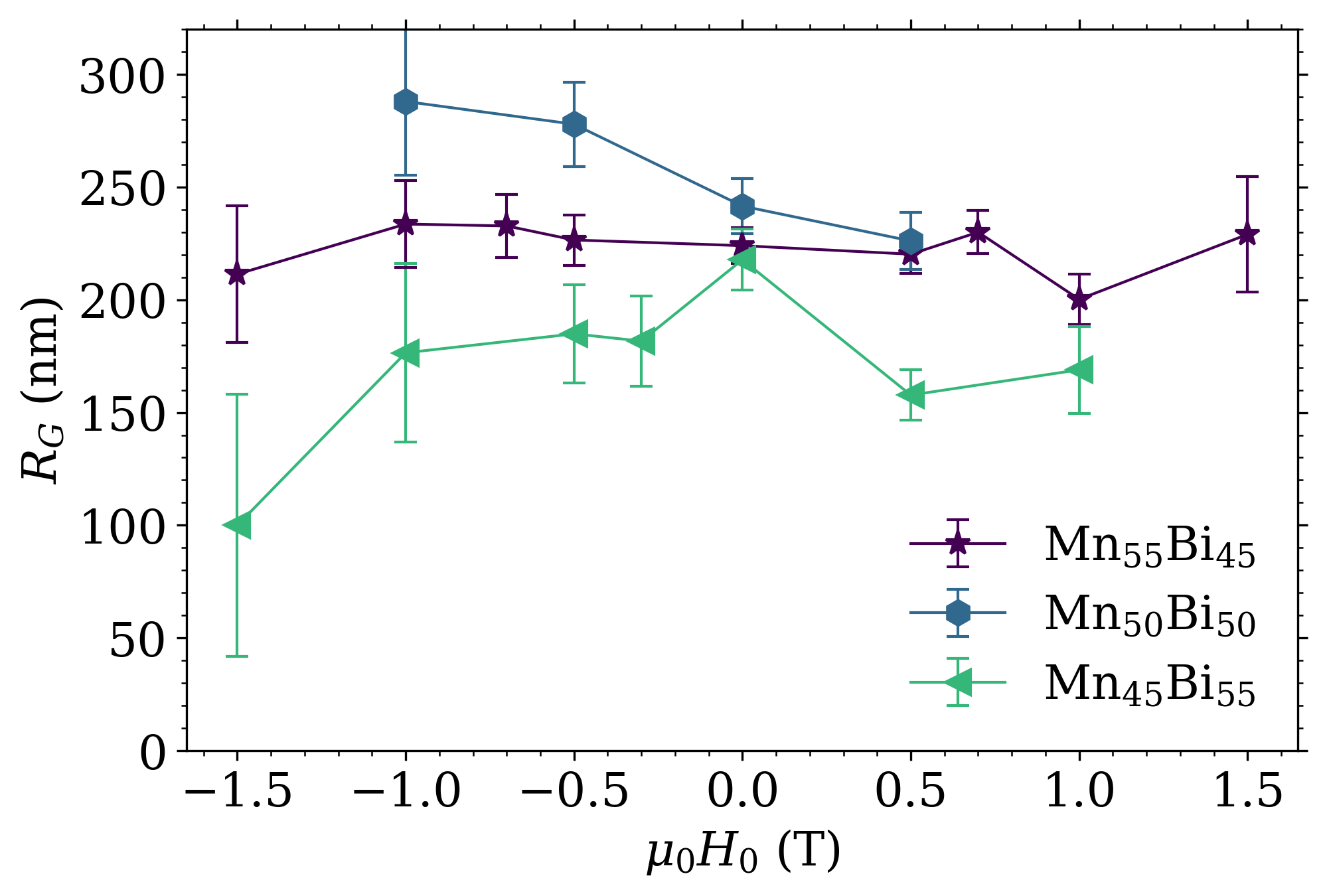}
\caption{Magnetic field dependence of the Guinier radii $R_G$ resulting from the generalized Guinier-Porod model. Lines are a guide to the eyes.}
\label{rgfielddep}
\end{figure}

In order to quantify the range of the magnetic correlations we have azimuthally-averaged the two-dimensional magnetic SANS cross sections and fitted the resulting data to the generalized Guinier-Porod (GP) model [Eqs.~(\ref{gpmodeleq1})$-$(\ref{gpmodeleq2})]. The results of the weighted nonlinear least-squares fitting procedure for the remanent-state data are displayed in Fig.~\ref{1dguinierfit} (solid lines) and demonstrate that the GP~model can very well describe the $q$-dependence of $d \Sigma_M / d \Omega$~\cite{suppmaterial}. The obtained Guinier radii $R_G$ are shown in Fig.~\ref{rgfielddep}, while Table~\ref{tab1} lists (for the remanent state) the results for the remaining fit parameters, the dimensionality parameter $s$ and the asymptotic power-law exponent $n$.

The origin of magnetic SANS is due to spatial mesoscale variations in the magnitude and orientation of the magnetization. Such magnetization fluctuations may be caused by microstructural defects (e.g., dislocations, interfaces, pores) via the magnetoelastic coupling of the magnetization to the strain field of the defect~\cite{kronfahn03}. The range and the amplitude of defect-induced spin disorder can be suppressed by an applied field. In the following we associate the value of $R_G$ with the size of such perturbed, nonuniformly magnetized regions around defects.

\begin{table}[tb!]
\caption{\label{tab1} Results of the fit analysis on Mn--Bi rare-earth-free permanent magnets using the generalized Guinier-Porod model~\cite{hammouda2010a} (remanent state).}
\begin{ruledtabular}
\begin{tabular}{cccc}
 & Mn$_{55}$Bi$_{45}$ & Mn$_{50}$Bi$_{50}$ & Mn$_{45}$Bi$_{55}$ \\ \hline
$R_G$~($\mathrm{nm}$) & 224 $\pm$   8 & 242 $\pm$  12 & 218 $\pm$  13 \\
$s$ & 0.30 $\pm$ 0.07 & 0.35 $\pm$ 0.10 & 1.07 $\pm$ 0.09 \\
$n$ & 3.38 $\pm$ 0.04 & 3.42 $\pm$ 0.04 & 3.58 $\pm$ 0.03 
\end{tabular}
\end{ruledtabular}
\end{table}

The Guinier radii in Fig.~\ref{rgfielddep} do not exhibit a systematic variation with the composition of the Mn--Bi samples. At remanence, their values range between $R_G \sim 220$$-$$240 \, \mathrm{nm}$. While the $R_G$ for the Mn$_{55}$Bi$_{45}$ sample are field independent within error bars, the Mn$_{45}$Bi$_{55}$ specimen exhibits a decrease of $R_G$ with increasing field, from about $220 \, \mathrm{nm}$ at remanence to $\sim 100  \, \mathrm{nm}$ at $1.5  \, \mathrm{T}$. Such a behavior is in qualitative agreement with the suppression of spin-misalignment fluctuations around defects with increasing applied field~\cite{Mettus2015}. On the other hand, the Mn$_{50}$Bi$_{50}$ specimen seems to exhibit an increase of $R_G$ with increasing field, from about $240 \, \mathrm{nm}$ at remanence to $\sim 285  \, \mathrm{nm}$ at $1.0  \, \mathrm{T}$. However, in view of the large uncertainties in the $R_G$-values of this sample, an unambiguous determination of the field behavior of the $R_G(H_0)$~data set is difficult.

The Porod exponents of $n \sim 3.4$$-$$3.6$ are systematically reduced below the sharp-interface value of $n=4$. In the context of particle scattering this observation could be interpreted as a smoothing of the surfaces of the scattering objects~\cite{hammouda2010a}. However, for magnetic SANS, where continuous rather than sharp scattering-length density variations are at the origin of the scattering, asymptotic power-law exponents smaller than 4 have only been reported for amorphous magnets~\cite{mettusprm2017}. Similarly, exponentially correlated magnetization fluctuations would give rise to $n=4$, corresponding to a Lorentzian-squared cross section. Therefore, the unusually low $n$-values observed in Mn--Bi remain to be explored by future experimental and theoretical neutron studies. 

Within the generalized Guinier-Porod model the $s$-parameter models nonspherical objects~\cite{hammouda2010a}. For three-dimensional globular particles (or domains), $s$ is expected to take on a value of $s=0$. The Mn$_{55}$Bi$_{45}$ and Mn$_{50}$Bi$_{50}$ samples are close to this value, whereas Mn$_{45}$Bi$_{55}$ exhibits $s=1.07$, which would indicate scattering due to elongated rod-like objects. The latter observation is surprising in view of the fact that extended electron-microscopy investigations on similar samples, albeit on a different length scale, did not reveal the presence of shape-anisotropic particles~\cite{Chen2016}.

\begin{figure}[tb!]
\centering
\includegraphics[width=1.0\linewidth]{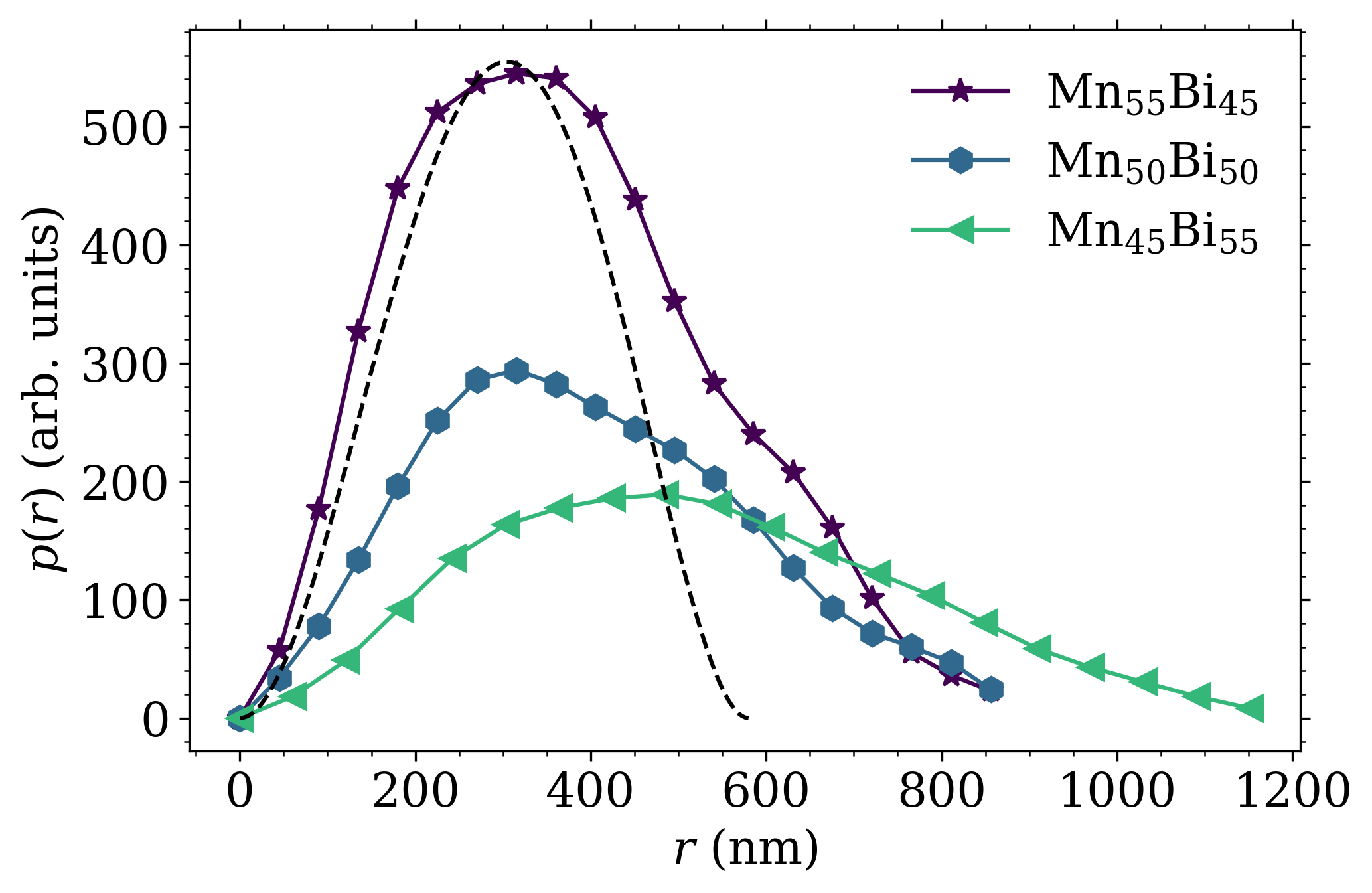}
\caption{Distance distribution functions $p(r)$ [Eq.~(\ref{pvonrfunc})] of the remanent-state Mn--Bi data shown in Fig.~\ref{1dguinierfit}. Dashed line:~Analytical $p(r) \propto r^2 (1 - \frac{3r}{4R} + \frac{r^3}{16R^3})$ of a sphere of radius $R = 290 \, \mathrm{nm}$, corresponding to a Guinier radius of $R_G = \sqrt{\frac{3}{5}} R = 225 \, \mathrm{nm}$.}
\label{pRfits}
\end{figure}

In order to further understand the differences between the samples (regarding the $s$-parameter), we have model-independently calculated the distance distribution function $p(r)$ [Eq.~(\ref{pvonrfunc})]. The results for $p(r)$ in Fig.~\ref{pRfits} are qualitatively consistent with the numerical fit analysis using the generalized Guinier-Porod model. The Mn$_{55}$Bi$_{45}$ and Mn$_{50}$Bi$_{50}$ samples both exhibit a $p(r)$ which is typical for globular scatterers~\cite{rgcomment}. Yet, a small shoulder at the larger distances points towards the presence of slightly anisotropic structures. By contrast, the $p(r)$ of the Mn$_{45}$Bi$_{55}$ sample clearly shows a broad maximum at $r \cong 470 \, \mathrm{nm}$ followed by a long tail at the larger $r$, suggesting that the scattering originates from shape-anisotropic elongated objects (compare Fig.~5 in the review by \textcite{svergun03}). The broad maximum of $p(r)$ at the smaller distances of the Mn$_{45}$Bi$_{55}$ specimen corresponds to the shorter dimension of the structure. This finding is in line with the behavior of the $s$-parameter obtained from the Guinier-Porod model.

The Guinier radius $R_G$, which is one of the central outcomes of our neutron analysis (Fig.~\ref{rgfielddep}), represents the characteristic size over which microstructural-defect-induced perturbations in the spin structure are transmitted by the exchange interaction into the surrounding crystal lattice; in other words, $R_G$ is considered to be a measure for the size of inhomogeneously magnetized regions around lattice imperfections. This length scale is of relevance for the understanding of the coercivity mechanism in Mn--Bi magnets---domain nucleation versus pinning---which is currently discussed in the literature~\cite{Curcio2015,Muralidhar2017,Zamora2018}. For instance, the nucleation of a reverse domain in a grain usually starts at a defect site, where the magnetic anisotropy may be reduced relative to the bulk phase. Therefore, the presented neutron methodology (analysis of difference data using the generalized Guinier-Porod model and calculation of the distance ditribution function) provides a means to systematically correlate the spin-misalignment length, which is a property of the defect, to the macroscopic parameters (e.g., coercivity, maximum energy product) of a permanent magnet. Moreover, previous studies (e.g., \cite{Muralidhar2017}) demonstrated enhanced coercivity over a wide temperature range with shifting alloy composition towards Bi, which was explained by differences in the grain-size distribution. Our SANS analysis indicates that an increase of the Bi content results in increasingly elongated magnetic structures (Fig.~\ref{pRfits}). Thus, a further increase of Bi might be a valid approach to enhance the magnetic hardness of the compound via shape anisotropy. In this respect, magnetic SANS permits the determination of the relevant figures of merit ($R_G$, $s$, $n$), which are otherwise not accessible by integral measurement techniques.

\section{Conclusion and outlook}
\label{conclusion}

We have investigated the magnetic microstructure of rare-earth-free Mn--Bi magnets by means of unpolarized very small-angle neutron scattering (SANS). The magnetic scattering cross section, which has been obtained by subtracting the total nuclear and magnetic scattering signal at $2.2.\, \mathrm{T}$ from data at lower fields, has been described in terms of the generalized Guinier-Porod model. The value of the Guinier radius is interpreted as the size of inhomogeneously magnetized regions around microstructural defects. We find that the spin-misalignment correlations are in the range of $\sim 100$$-$$300 \, \mathrm{nm}$ for the compositions studied. Moreover, in particular using the distance distribution function, our analysis indicates that the magnetic scattering of the Mn$_{45}$Bi$_{55}$ sample is related to shape-anisotropic structures, while the scattering of Mn$_{55}$Bi$_{45}$ and Mn$_{50}$Bi$_{50}$ has its origin in more globular-like objects. The neutron-data subtraction procedure (low field minus high field) eliminates the nuclear scattering contribution, which is not further analyzed. In this respect, neutron imaging techniques could be employed for the characterization of the nuclear grain microstructure and morphology inside the bulk of the magnet~\cite{WROBLEWSKI1999}. In future investigations the usage of polarized neutrons will be beneficial (e.g., \cite{ibarra2006,bender2018prb,krycka2019,michelsheusler2019}), since it then becomes possible to directly measure the purely magnetic SANS cross section without the coherent nuclear contribution. Likewise, extending the range of momentum transfers to the so-called ultra SANS regime ($q_{\mathrm{min}} \cong 10^{-4} \, \mathrm{nm}^{-1}$) permits following the correlations up to the $10$~micron range. Temperature-dependent neutron measurements will allow one to obtain mesoscale information on the relation between the magnetic microstructure and the positive temperature coefficient of the magnetic anisotropy found for this material.

\section*{Acknowledgements}

Artem Malyeyev, Philipp Bender, and Andreas Michels acknowledge financial support from the National Research Fund of Luxembourg (AFR and CORE SANS4NCC grants). We thank the Heinz Maier-Leibnitz Zentrum for the provision of neutron beamtime.


\begin{thebibliography}{74}%
\makeatletter
\providecommand \@ifxundefined [1]{%
 \@ifx{#1\undefined}
}%
\providecommand \@ifnum [1]{%
 \ifnum #1\expandafter \@firstoftwo
 \else \expandafter \@secondoftwo
 \fi
}%
\providecommand \@ifx [1]{%
 \ifx #1\expandafter \@firstoftwo
 \else \expandafter \@secondoftwo
 \fi
}%
\providecommand \natexlab [1]{#1}%
\providecommand \enquote  [1]{``#1''}%
\providecommand \bibnamefont  [1]{#1}%
\providecommand \bibfnamefont [1]{#1}%
\providecommand \citenamefont [1]{#1}%
\providecommand \href@noop [0]{\@secondoftwo}%
\providecommand \href [0]{\begingroup \@sanitize@url \@href}%
\providecommand \@href[1]{\@@startlink{#1}\@@href}%
\providecommand \@@href[1]{\endgroup#1\@@endlink}%
\providecommand \@sanitize@url [0]{\catcode `\\12\catcode `\$12\catcode
  `\&12\catcode `\#12\catcode `\^12\catcode `\_12\catcode `\%12\relax}%
\providecommand \@@startlink[1]{}%
\providecommand \@@endlink[0]{}%
\providecommand \url  [0]{\begingroup\@sanitize@url \@url }%
\providecommand \@url [1]{\endgroup\@href {#1}{\urlprefix }}%
\providecommand \urlprefix  [0]{URL }%
\providecommand \Eprint [0]{\href }%
\providecommand \doibase [0]{https://doi.org/}%
\providecommand \selectlanguage [0]{\@gobble}%
\providecommand \bibinfo  [0]{\@secondoftwo}%
\providecommand \bibfield  [0]{\@secondoftwo}%
\providecommand \translation [1]{[#1]}%
\providecommand \BibitemOpen [0]{}%
\providecommand \bibitemStop [0]{}%
\providecommand \bibitemNoStop [0]{.\EOS\space}%
\providecommand \EOS [0]{\spacefactor3000\relax}%
\providecommand \BibitemShut  [1]{\csname bibitem#1\endcsname}%
\let\auto@bib@innerbib\@empty
\bibitem [{\citenamefont {Gutfleisch}\ \emph {et~al.}(2011)\citenamefont
  {Gutfleisch}, \citenamefont {Willard}, \citenamefont {Br\"uck}, \citenamefont
  {Chen}, \citenamefont {Sankar},\ and\ \citenamefont {Liu}}]{Gutfleisch2011}%
  \BibitemOpen
  \bibfield  {author} {\bibinfo {author} {\bibfnamefont {O.}~\bibnamefont
  {Gutfleisch}}, \bibinfo {author} {\bibfnamefont {M.~A.}\ \bibnamefont
  {Willard}}, \bibinfo {author} {\bibfnamefont {E.}~\bibnamefont {Br\"uck}},
  \bibinfo {author} {\bibfnamefont {C.~H.}\ \bibnamefont {Chen}}, \bibinfo
  {author} {\bibfnamefont {S.~G.}\ \bibnamefont {Sankar}},\ and\ \bibinfo
  {author} {\bibfnamefont {J.~P.}\ \bibnamefont {Liu}},\ }\href@noop {}
  {\bibfield  {journal} {\bibinfo  {journal} {Adv. Mater.}\ }\textbf {\bibinfo
  {volume} {23}},\ \bibinfo {pages} {821} (\bibinfo {year} {2011})}\BibitemShut
  {NoStop}%
\bibitem [{\citenamefont {Riba}\ \emph {et~al.}(2016)\citenamefont {Riba},
  \citenamefont {L{\'{o}}pez-Torres}, \citenamefont {Romeral},\ and\
  \citenamefont {Garcia}}]{Riba2016}%
  \BibitemOpen
  \bibfield  {author} {\bibinfo {author} {\bibfnamefont {J.~R.}\ \bibnamefont
  {Riba}}, \bibinfo {author} {\bibfnamefont {C.}~\bibnamefont
  {L{\'{o}}pez-Torres}}, \bibinfo {author} {\bibfnamefont {L.}~\bibnamefont
  {Romeral}},\ and\ \bibinfo {author} {\bibfnamefont {A.}~\bibnamefont
  {Garcia}},\ }\href {https://doi.org/10.1016/j.rser.2015.12.121} {\bibinfo
  {title} {{Rare-earth-free propulsion motors for electric vehicles: A
  technology review}}} (\bibinfo {year} {2016})\BibitemShut {NoStop}%
\bibitem [{\citenamefont {Coey}(2012)}]{Coey2012}%
  \BibitemOpen
  \bibfield  {author} {\bibinfo {author} {\bibfnamefont {J.}~\bibnamefont
  {Coey}},\ }\href {https://doi.org/10.1016/j.scriptamat.2012.04.036}
  {\bibfield  {journal} {\bibinfo  {journal} {Scripta Materialia}\ }\textbf
  {\bibinfo {volume} {67}},\ \bibinfo {pages} {524} (\bibinfo {year}
  {2012})}\BibitemShut {NoStop}%
\bibitem [{\citenamefont {Coey}(2014)}]{Coey2014}%
  \BibitemOpen
  \bibfield  {author} {\bibinfo {author} {\bibfnamefont {J.~M.~D.}\
  \bibnamefont {Coey}},\ }\href@noop {} {\bibfield  {journal} {\bibinfo
  {journal} {J. Phys.: Condens. Matter}\ }\textbf {\bibinfo {volume} {26}},\
  \bibinfo {pages} {064211} (\bibinfo {year} {2014})}\BibitemShut {NoStop}%
\bibitem [{\citenamefont {Ener}\ \emph {et~al.}(2015)\citenamefont {Ener},
  \citenamefont {Skokov}, \citenamefont {Karpenkov}, \citenamefont {Kuz'min},\
  and\ \citenamefont {Gutfleisch}}]{Ener2015}%
  \BibitemOpen
  \bibfield  {author} {\bibinfo {author} {\bibfnamefont {S.}~\bibnamefont
  {Ener}}, \bibinfo {author} {\bibfnamefont {K.~P.}\ \bibnamefont {Skokov}},
  \bibinfo {author} {\bibfnamefont {D.~Y.}\ \bibnamefont {Karpenkov}}, \bibinfo
  {author} {\bibfnamefont {M.~D.}\ \bibnamefont {Kuz'min}},\ and\ \bibinfo
  {author} {\bibfnamefont {O.}~\bibnamefont {Gutfleisch}},\ }\href
  {https://doi.org/10.1016/j.jmmm.2015.02.001} {\bibfield  {journal} {\bibinfo
  {journal} {J. Magn. Magn. Mater.}\ }\textbf {\bibinfo {volume} {382}},\
  \bibinfo {pages} {265} (\bibinfo {year} {2015})}\BibitemShut {NoStop}%
\bibitem [{\citenamefont {Jian}\ \emph {et~al.}(2015)\citenamefont {Jian},
  \citenamefont {Skokov},\ and\ \citenamefont {Gutfleisch}}]{Jian2015}%
  \BibitemOpen
  \bibfield  {author} {\bibinfo {author} {\bibfnamefont {H.}~\bibnamefont
  {Jian}}, \bibinfo {author} {\bibfnamefont {K.~P.}\ \bibnamefont {Skokov}},\
  and\ \bibinfo {author} {\bibfnamefont {O.}~\bibnamefont {Gutfleisch}},\
  }\href {https://doi.org/10.1016/j.jallcom.2014.10.138} {\bibfield  {journal}
  {\bibinfo  {journal} {J. Alloys Compd.}\ }\textbf {\bibinfo {volume} {622}},\
  \bibinfo {pages} {524} (\bibinfo {year} {2015})}\BibitemShut {NoStop}%
\bibitem [{\citenamefont {Jia}\ \emph {et~al.}(2020)\citenamefont {Jia},
  \citenamefont {Wu}, \citenamefont {Zhao}, \citenamefont {Zuo}, \citenamefont
  {Skokov}, \citenamefont {Gutfleisch}, \citenamefont {Jiang},\ and\
  \citenamefont {Xu}}]{gutfleisch2020}%
  \BibitemOpen
  \bibfield  {author} {\bibinfo {author} {\bibfnamefont {Y.}~\bibnamefont
  {Jia}}, \bibinfo {author} {\bibfnamefont {Y.}~\bibnamefont {Wu}}, \bibinfo
  {author} {\bibfnamefont {S.}~\bibnamefont {Zhao}}, \bibinfo {author}
  {\bibfnamefont {S.}~\bibnamefont {Zuo}}, \bibinfo {author} {\bibfnamefont
  {K.~P.}\ \bibnamefont {Skokov}}, \bibinfo {author} {\bibfnamefont
  {O.}~\bibnamefont {Gutfleisch}}, \bibinfo {author} {\bibfnamefont
  {C.}~\bibnamefont {Jiang}},\ and\ \bibinfo {author} {\bibfnamefont
  {H.}~\bibnamefont {Xu}},\ }\href
  {https://doi.org/10.1103/PhysRevMaterials.4.094402} {\bibfield  {journal}
  {\bibinfo  {journal} {Phys. Rev. Materials}\ }\textbf {\bibinfo {volume}
  {4}},\ \bibinfo {pages} {094402} (\bibinfo {year} {2020})}\BibitemShut
  {NoStop}%
\bibitem [{\citenamefont {Park}\ \emph {et~al.}(2014)\citenamefont {Park},
  \citenamefont {Hong}, \citenamefont {Lee}, \citenamefont {Lee}, \citenamefont
  {Kim},\ and\ \citenamefont {Choi}}]{Park2014}%
  \BibitemOpen
  \bibfield  {author} {\bibinfo {author} {\bibfnamefont {J.}~\bibnamefont
  {Park}}, \bibinfo {author} {\bibfnamefont {Y.-K.}\ \bibnamefont {Hong}},
  \bibinfo {author} {\bibfnamefont {J.}~\bibnamefont {Lee}}, \bibinfo {author}
  {\bibfnamefont {W.}~\bibnamefont {Lee}}, \bibinfo {author} {\bibfnamefont
  {S.-G.}\ \bibnamefont {Kim}},\ and\ \bibinfo {author} {\bibfnamefont {C.-J.}\
  \bibnamefont {Choi}},\ }\href {https://doi.org/10.3390/met4030455} {\bibfield
   {journal} {\bibinfo  {journal} {Metals}\ }\textbf {\bibinfo {volume} {4}},\
  \bibinfo {pages} {455} (\bibinfo {year} {2014})}\BibitemShut {NoStop}%
\bibitem [{\citenamefont {Ly}\ \emph {et~al.}(2015)\citenamefont {Ly},
  \citenamefont {Wu}, \citenamefont {Smillie}, \citenamefont {Shoji},
  \citenamefont {Kato}, \citenamefont {Manabe},\ and\ \citenamefont
  {Suzuki}}]{Ly2014}%
  \BibitemOpen
  \bibfield  {author} {\bibinfo {author} {\bibfnamefont {V.}~\bibnamefont
  {Ly}}, \bibinfo {author} {\bibfnamefont {X.}~\bibnamefont {Wu}}, \bibinfo
  {author} {\bibfnamefont {L.}~\bibnamefont {Smillie}}, \bibinfo {author}
  {\bibfnamefont {T.}~\bibnamefont {Shoji}}, \bibinfo {author} {\bibfnamefont
  {A.}~\bibnamefont {Kato}}, \bibinfo {author} {\bibfnamefont {A.}~\bibnamefont
  {Manabe}},\ and\ \bibinfo {author} {\bibfnamefont {K.}~\bibnamefont
  {Suzuki}},\ }\href {https://doi.org/10.1016/j.jallcom.2014.01.120} {\bibfield
   {journal} {\bibinfo  {journal} {Journal of Alloys and Compounds}\ }\textbf
  {\bibinfo {volume} {615}},\ \bibinfo {pages} {S285} (\bibinfo {year}
  {2015})}\BibitemShut {NoStop}%
\bibitem [{\citenamefont {Baker}(2015)}]{Baker2015}%
  \BibitemOpen
  \bibfield  {author} {\bibinfo {author} {\bibfnamefont {I.}~\bibnamefont
  {Baker}},\ }\href {https://doi.org/10.3390/met5031435} {\bibfield  {journal}
  {\bibinfo  {journal} {Metals}\ }\textbf {\bibinfo {volume} {5}},\ \bibinfo
  {pages} {1435} (\bibinfo {year} {2015})}\BibitemShut {NoStop}%
\bibitem [{\citenamefont {Chen}\ \emph {et~al.}(2015)\citenamefont {Chen},
  \citenamefont {Gregori}, \citenamefont {Leineweber}, \citenamefont {Qu},
  \citenamefont {Chen}, \citenamefont {Tietze}, \citenamefont
  {Kronm{\"{u}}ller}, \citenamefont {Sch{\"{u}}tz},\ and\ \citenamefont
  {Goering}}]{Chen2015}%
  \BibitemOpen
  \bibfield  {author} {\bibinfo {author} {\bibfnamefont {Y.~C.}\ \bibnamefont
  {Chen}}, \bibinfo {author} {\bibfnamefont {G.}~\bibnamefont {Gregori}},
  \bibinfo {author} {\bibfnamefont {A.}~\bibnamefont {Leineweber}}, \bibinfo
  {author} {\bibfnamefont {F.}~\bibnamefont {Qu}}, \bibinfo {author}
  {\bibfnamefont {C.~C.}\ \bibnamefont {Chen}}, \bibinfo {author}
  {\bibfnamefont {T.}~\bibnamefont {Tietze}}, \bibinfo {author} {\bibfnamefont
  {H.}~\bibnamefont {Kronm{\"{u}}ller}}, \bibinfo {author} {\bibfnamefont
  {G.}~\bibnamefont {Sch{\"{u}}tz}},\ and\ \bibinfo {author} {\bibfnamefont
  {E.}~\bibnamefont {Goering}},\ }\href
  {https://doi.org/10.1016/j.scriptamat.2015.06.003} {\bibfield  {journal}
  {\bibinfo  {journal} {Scripta Materialia}\ }\textbf {\bibinfo {volume}
  {107}},\ \bibinfo {pages} {131} (\bibinfo {year} {2015})}\BibitemShut
  {NoStop}%
\bibitem [{\citenamefont {Kim}\ \emph {et~al.}(2017)\citenamefont {Kim},
  \citenamefont {Moon}, \citenamefont {Jung}, \citenamefont {Kim},
  \citenamefont {Lee}, \citenamefont {Choi-Yim},\ and\ \citenamefont
  {Lee}}]{Kim2017}%
  \BibitemOpen
  \bibfield  {author} {\bibinfo {author} {\bibfnamefont {S.-M.}\ \bibnamefont
  {Kim}}, \bibinfo {author} {\bibfnamefont {H.}~\bibnamefont {Moon}}, \bibinfo
  {author} {\bibfnamefont {H.}~\bibnamefont {Jung}}, \bibinfo {author}
  {\bibfnamefont {S.-M.}\ \bibnamefont {Kim}}, \bibinfo {author} {\bibfnamefont
  {H.-S.}\ \bibnamefont {Lee}}, \bibinfo {author} {\bibfnamefont
  {H.}~\bibnamefont {Choi-Yim}},\ and\ \bibinfo {author} {\bibfnamefont
  {W.}~\bibnamefont {Lee}},\ }\href
  {https://doi.org/10.1016/j.jallcom.2017.03.067} {\bibfield  {journal}
  {\bibinfo  {journal} {Journal of Alloys and Compounds}\ }\textbf {\bibinfo
  {volume} {708}},\ \bibinfo {pages} {1245} (\bibinfo {year}
  {2017})}\BibitemShut {NoStop}%
\bibitem [{\citenamefont {Nguyen}\ and\ \citenamefont
  {Nguyen}(2018)}]{Nguyen2018}%
  \BibitemOpen
  \bibfield  {author} {\bibinfo {author} {\bibfnamefont {V.~V.}\ \bibnamefont
  {Nguyen}}\ and\ \bibinfo {author} {\bibfnamefont {T.~X.}\ \bibnamefont
  {Nguyen}},\ }\href {https://doi.org/10.1016/j.physb.2017.06.018} {\bibfield
  {journal} {\bibinfo  {journal} {Physica B: Condensed Matter}\ }\textbf
  {\bibinfo {volume} {532}},\ \bibinfo {pages} {103} (\bibinfo {year}
  {2018})}\BibitemShut {NoStop}%
\bibitem [{\citenamefont {Nguyen}\ \emph {et~al.}(2014)\citenamefont {Nguyen},
  \citenamefont {Poudyal}, \citenamefont {Liu}, \citenamefont {Liu},
  \citenamefont {Sun}, \citenamefont {Kramer},\ and\ \citenamefont
  {Cui}}]{Nguyen2014}%
  \BibitemOpen
  \bibfield  {author} {\bibinfo {author} {\bibfnamefont {V.~V.}\ \bibnamefont
  {Nguyen}}, \bibinfo {author} {\bibfnamefont {N.}~\bibnamefont {Poudyal}},
  \bibinfo {author} {\bibfnamefont {X.~B.}\ \bibnamefont {Liu}}, \bibinfo
  {author} {\bibfnamefont {J.~P.}\ \bibnamefont {Liu}}, \bibinfo {author}
  {\bibfnamefont {K.}~\bibnamefont {Sun}}, \bibinfo {author} {\bibfnamefont
  {M.~J.}\ \bibnamefont {Kramer}},\ and\ \bibinfo {author} {\bibfnamefont
  {J.}~\bibnamefont {Cui}},\ }\href
  {https://doi.org/10.1088/2053-1591/1/3/036108} {\bibfield  {journal}
  {\bibinfo  {journal} {Materials Research Express}\ }\textbf {\bibinfo
  {volume} {1}},\ \bibinfo {pages} {036108} (\bibinfo {year}
  {2014})}\BibitemShut {NoStop}%
\bibitem [{\citenamefont {Cui}\ \emph {et~al.}(2014)\citenamefont {Cui},
  \citenamefont {Choi}, \citenamefont {Polikarpov}, \citenamefont {Bowden},
  \citenamefont {Xie}, \citenamefont {Li}, \citenamefont {Nie}, \citenamefont
  {Zarkevich}, \citenamefont {Kramer},\ and\ \citenamefont
  {Johnson}}]{Cui2014b}%
  \BibitemOpen
  \bibfield  {author} {\bibinfo {author} {\bibfnamefont {J.}~\bibnamefont
  {Cui}}, \bibinfo {author} {\bibfnamefont {J.~P.}\ \bibnamefont {Choi}},
  \bibinfo {author} {\bibfnamefont {E.}~\bibnamefont {Polikarpov}}, \bibinfo
  {author} {\bibfnamefont {M.~E.}\ \bibnamefont {Bowden}}, \bibinfo {author}
  {\bibfnamefont {W.}~\bibnamefont {Xie}}, \bibinfo {author} {\bibfnamefont
  {G.}~\bibnamefont {Li}}, \bibinfo {author} {\bibfnamefont {Z.}~\bibnamefont
  {Nie}}, \bibinfo {author} {\bibfnamefont {N.}~\bibnamefont {Zarkevich}},
  \bibinfo {author} {\bibfnamefont {M.~J.}\ \bibnamefont {Kramer}},\ and\
  \bibinfo {author} {\bibfnamefont {D.}~\bibnamefont {Johnson}},\ }\href
  {https://doi.org/10.1016/j.actamat.2014.07.034} {\bibfield  {journal}
  {\bibinfo  {journal} {Acta Materialia}\ }\textbf {\bibinfo {volume} {79}},\
  \bibinfo {pages} {374} (\bibinfo {year} {2014})}\BibitemShut {NoStop}%
\bibitem [{\citenamefont {Poudyal}\ \emph {et~al.}(2016)\citenamefont
  {Poudyal}, \citenamefont {Liu}, \citenamefont {Wang}, \citenamefont {Nguyen},
  \citenamefont {Ma}, \citenamefont {Gandha}, \citenamefont {Elkins},
  \citenamefont {Liu}, \citenamefont {Sun}, \citenamefont {Kramer},\ and\
  \citenamefont {Cui}}]{Poudyal2016}%
  \BibitemOpen
  \bibfield  {author} {\bibinfo {author} {\bibfnamefont {N.}~\bibnamefont
  {Poudyal}}, \bibinfo {author} {\bibfnamefont {X.}~\bibnamefont {Liu}},
  \bibinfo {author} {\bibfnamefont {W.}~\bibnamefont {Wang}}, \bibinfo {author}
  {\bibfnamefont {V.~V.}\ \bibnamefont {Nguyen}}, \bibinfo {author}
  {\bibfnamefont {Y.}~\bibnamefont {Ma}}, \bibinfo {author} {\bibfnamefont
  {K.}~\bibnamefont {Gandha}}, \bibinfo {author} {\bibfnamefont
  {K.}~\bibnamefont {Elkins}}, \bibinfo {author} {\bibfnamefont {J.~P.}\
  \bibnamefont {Liu}}, \bibinfo {author} {\bibfnamefont {K.}~\bibnamefont
  {Sun}}, \bibinfo {author} {\bibfnamefont {M.~J.}\ \bibnamefont {Kramer}},\
  and\ \bibinfo {author} {\bibfnamefont {J.}~\bibnamefont {Cui}},\ }\href
  {https://doi.org/10.1063/1.4942955} {\bibfield  {journal} {\bibinfo
  {journal} {AIP Advances}\ }\textbf {\bibinfo {volume} {6}},\ \bibinfo {pages}
  {056004} (\bibinfo {year} {2016})}\BibitemShut {NoStop}%
\bibitem [{\citenamefont {Mitsui}\ \emph {et~al.}(2016)\citenamefont {Mitsui},
  \citenamefont {Abematsu}, \citenamefont {Umetsu}, \citenamefont {Takahashi},\
  and\ \citenamefont {Koyama}}]{Mitsui2016}%
  \BibitemOpen
  \bibfield  {author} {\bibinfo {author} {\bibfnamefont {Y.}~\bibnamefont
  {Mitsui}}, \bibinfo {author} {\bibfnamefont {K.~I.}\ \bibnamefont
  {Abematsu}}, \bibinfo {author} {\bibfnamefont {R.~Y.}\ \bibnamefont
  {Umetsu}}, \bibinfo {author} {\bibfnamefont {K.}~\bibnamefont {Takahashi}},\
  and\ \bibinfo {author} {\bibfnamefont {K.}~\bibnamefont {Koyama}},\ }\href
  {https://doi.org/10.1016/j.jmmm.2015.07.114} {\bibfield  {journal} {\bibinfo
  {journal} {Journal of Magnetism and Magnetic Materials}\ }\textbf {\bibinfo
  {volume} {400}},\ \bibinfo {pages} {304} (\bibinfo {year}
  {2016})}\BibitemShut {NoStop}%
\bibitem [{\citenamefont {Xiang}\ \emph
  {et~al.}(2018{\natexlab{a}})\citenamefont {Xiang}, \citenamefont {Song},
  \citenamefont {Pan}, \citenamefont {Shen}, \citenamefont {Qian},
  \citenamefont {Luo}, \citenamefont {Liu}, \citenamefont {Yang}, \citenamefont
  {Yan},\ and\ \citenamefont {Lu}}]{Xiang2018}%
  \BibitemOpen
  \bibfield  {author} {\bibinfo {author} {\bibfnamefont {Z.}~\bibnamefont
  {Xiang}}, \bibinfo {author} {\bibfnamefont {Y.}~\bibnamefont {Song}},
  \bibinfo {author} {\bibfnamefont {D.}~\bibnamefont {Pan}}, \bibinfo {author}
  {\bibfnamefont {Y.}~\bibnamefont {Shen}}, \bibinfo {author} {\bibfnamefont
  {L.}~\bibnamefont {Qian}}, \bibinfo {author} {\bibfnamefont {Z.}~\bibnamefont
  {Luo}}, \bibinfo {author} {\bibfnamefont {Y.}~\bibnamefont {Liu}}, \bibinfo
  {author} {\bibfnamefont {H.}~\bibnamefont {Yang}}, \bibinfo {author}
  {\bibfnamefont {H.}~\bibnamefont {Yan}},\ and\ \bibinfo {author}
  {\bibfnamefont {W.}~\bibnamefont {Lu}},\ }\href
  {https://doi.org/10.1016/j.jallcom.2018.02.102} {\bibfield  {journal}
  {\bibinfo  {journal} {Journal of Alloys and Compounds}\ }\textbf {\bibinfo
  {volume} {744}},\ \bibinfo {pages} {432} (\bibinfo {year}
  {2018}{\natexlab{a}})}\BibitemShut {NoStop}%
\bibitem [{\citenamefont {Xiang}\ \emph
  {et~al.}(2018{\natexlab{b}})\citenamefont {Xiang}, \citenamefont {Xu},
  \citenamefont {Wang}, \citenamefont {Song}, \citenamefont {Yang},\ and\
  \citenamefont {Lu}}]{Xiang2018b}%
  \BibitemOpen
  \bibfield  {author} {\bibinfo {author} {\bibfnamefont {Z.}~\bibnamefont
  {Xiang}}, \bibinfo {author} {\bibfnamefont {C.}~\bibnamefont {Xu}}, \bibinfo
  {author} {\bibfnamefont {T.}~\bibnamefont {Wang}}, \bibinfo {author}
  {\bibfnamefont {Y.}~\bibnamefont {Song}}, \bibinfo {author} {\bibfnamefont
  {H.}~\bibnamefont {Yang}},\ and\ \bibinfo {author} {\bibfnamefont
  {W.}~\bibnamefont {Lu}},\ }\href
  {https://doi.org/10.1016/j.intermet.2018.07.003} {\bibfield  {journal}
  {\bibinfo  {journal} {Intermetallics}\ }\textbf {\bibinfo {volume} {101}},\
  \bibinfo {pages} {13} (\bibinfo {year} {2018}{\natexlab{b}})}\BibitemShut
  {NoStop}%
\bibitem [{\citenamefont {Janotov}\ \emph {et~al.}(2018)\citenamefont
  {Janotov}, \citenamefont {{Svec}}, \citenamefont {Svec}, \citenamefont
  {Matko}, \citenamefont {Jani~ckovi}, \citenamefont {Kunca}, \citenamefont
  {Marcin},\ and\ \citenamefont {Skorv~anek}}]{Janotov2018}%
  \BibitemOpen
  \bibfield  {author} {\bibinfo {author} {\bibfnamefont {I.}~\bibnamefont
  {Janotov}}, \bibinfo {author} {\bibfnamefont {P.}~\bibnamefont {{Svec}}},
  \bibinfo {author} {\bibfnamefont {P.}~\bibnamefont {Svec}}, \bibinfo {author}
  {\bibfnamefont {I.}~\bibnamefont {Matko}}, \bibinfo {author} {\bibfnamefont
  {D.}~\bibnamefont {Jani~ckovi}}, \bibinfo {author} {\bibfnamefont
  {B.}~\bibnamefont {Kunca}}, \bibinfo {author} {\bibfnamefont
  {J.}~\bibnamefont {Marcin}},\ and\ \bibinfo {author} {\bibfnamefont
  {I.}~\bibnamefont {Skorv~anek}},\ }\href
  {https://doi.org/10.1016/j.jallcom.2018.03.208} {\bibfield  {journal}
  {\bibinfo  {journal} {Journal of Alloys and Compounds}\ }\textbf {\bibinfo
  {volume} {749}},\ \bibinfo {pages} {128} (\bibinfo {year}
  {2018})}\BibitemShut {NoStop}%
\bibitem [{\citenamefont {Cao}\ \emph {et~al.}(2019)\citenamefont {Cao},
  \citenamefont {Huang}, \citenamefont {Hou}, \citenamefont {Shi},
  \citenamefont {Yan}, \citenamefont {Zhong},\ and\ \citenamefont
  {Wang}}]{Cao2019}%
  \BibitemOpen
  \bibfield  {author} {\bibinfo {author} {\bibfnamefont {J.}~\bibnamefont
  {Cao}}, \bibinfo {author} {\bibfnamefont {Y.~L.}\ \bibnamefont {Huang}},
  \bibinfo {author} {\bibfnamefont {Y.~H.}\ \bibnamefont {Hou}}, \bibinfo
  {author} {\bibfnamefont {Z.~Q.}\ \bibnamefont {Shi}}, \bibinfo {author}
  {\bibfnamefont {X.~T.}\ \bibnamefont {Yan}}, \bibinfo {author} {\bibfnamefont
  {Z.~C.}\ \bibnamefont {Zhong}},\ and\ \bibinfo {author} {\bibfnamefont
  {G.~P.}\ \bibnamefont {Wang}},\ }\href
  {https://doi.org/10.1016/j.jmmm.2018.10.052} {\bibfield  {journal} {\bibinfo
  {journal} {Journal of Magnetism and Magnetic Materials}\ }\textbf {\bibinfo
  {volume} {473}},\ \bibinfo {pages} {505} (\bibinfo {year}
  {2019})}\BibitemShut {NoStop}%
\bibitem [{\citenamefont {Disch}\ \emph {et~al.}(2012)\citenamefont {Disch},
  \citenamefont {Wetterskog}, \citenamefont {Hermann}, \citenamefont
  {Wiedenmann}, \citenamefont {Vainio}, \citenamefont {Salazar-Alvarez},
  \citenamefont {Bergstr\"om},\ and\ \citenamefont {Br\"uckel}}]{disch2012}%
  \BibitemOpen
  \bibfield  {author} {\bibinfo {author} {\bibfnamefont {S.}~\bibnamefont
  {Disch}}, \bibinfo {author} {\bibfnamefont {E.}~\bibnamefont {Wetterskog}},
  \bibinfo {author} {\bibfnamefont {R.~P.}\ \bibnamefont {Hermann}}, \bibinfo
  {author} {\bibfnamefont {A.}~\bibnamefont {Wiedenmann}}, \bibinfo {author}
  {\bibfnamefont {U.}~\bibnamefont {Vainio}}, \bibinfo {author} {\bibfnamefont
  {G.}~\bibnamefont {Salazar-Alvarez}}, \bibinfo {author} {\bibfnamefont
  {L.}~\bibnamefont {Bergstr\"om}},\ and\ \bibinfo {author} {\bibfnamefont
  {T.}~\bibnamefont {Br\"uckel}},\ }\href@noop {} {\bibfield  {journal}
  {\bibinfo  {journal} {New J. Phys.}\ }\textbf {\bibinfo {volume} {14}},\
  \bibinfo {pages} {013025} (\bibinfo {year} {2012})}\BibitemShut {NoStop}%
\bibitem [{\citenamefont {G\"unther}\ \emph {et~al.}(2014)\citenamefont
  {G\"unther}, \citenamefont {Honecker}, \citenamefont {Bick}, \citenamefont
  {Szary}, \citenamefont {Dewhurst}, \citenamefont {Keiderling}, \citenamefont
  {Feoktystov}, \citenamefont {Tsch\"ope}, \citenamefont {Birringer},\ and\
  \citenamefont {Michels}}]{guenther2014}%
  \BibitemOpen
  \bibfield  {author} {\bibinfo {author} {\bibfnamefont {A.}~\bibnamefont
  {G\"unther}}, \bibinfo {author} {\bibfnamefont {D.}~\bibnamefont {Honecker}},
  \bibinfo {author} {\bibfnamefont {J.-P.}\ \bibnamefont {Bick}}, \bibinfo
  {author} {\bibfnamefont {P.}~\bibnamefont {Szary}}, \bibinfo {author}
  {\bibfnamefont {C.~D.}\ \bibnamefont {Dewhurst}}, \bibinfo {author}
  {\bibfnamefont {U.}~\bibnamefont {Keiderling}}, \bibinfo {author}
  {\bibfnamefont {A.~V.}\ \bibnamefont {Feoktystov}}, \bibinfo {author}
  {\bibfnamefont {A.}~\bibnamefont {Tsch\"ope}}, \bibinfo {author}
  {\bibfnamefont {R.}~\bibnamefont {Birringer}},\ and\ \bibinfo {author}
  {\bibfnamefont {A.}~\bibnamefont {Michels}},\ }\href@noop {} {\bibfield
  {journal} {\bibinfo  {journal} {J. Appl. Cryst.}\ }\textbf {\bibinfo {volume}
  {47}},\ \bibinfo {pages} {992} (\bibinfo {year} {2014})}\BibitemShut
  {NoStop}%
\bibitem [{\citenamefont {Bender}\ \emph {et~al.}(2015)\citenamefont {Bender},
  \citenamefont {G\"unther}, \citenamefont {Honecker}, \citenamefont
  {Wiedenmann}, \citenamefont {Disch}, \citenamefont {Tsch\"ope}, \citenamefont
  {Michels},\ and\ \citenamefont {Birringer}}]{bender2015}%
  \BibitemOpen
  \bibfield  {author} {\bibinfo {author} {\bibfnamefont {P.}~\bibnamefont
  {Bender}}, \bibinfo {author} {\bibfnamefont {A.}~\bibnamefont {G\"unther}},
  \bibinfo {author} {\bibfnamefont {D.}~\bibnamefont {Honecker}}, \bibinfo
  {author} {\bibfnamefont {A.}~\bibnamefont {Wiedenmann}}, \bibinfo {author}
  {\bibfnamefont {S.}~\bibnamefont {Disch}}, \bibinfo {author} {\bibfnamefont
  {A.}~\bibnamefont {Tsch\"ope}}, \bibinfo {author} {\bibfnamefont
  {A.}~\bibnamefont {Michels}},\ and\ \bibinfo {author} {\bibfnamefont
  {R.}~\bibnamefont {Birringer}},\ }\href@noop {} {\bibfield  {journal}
  {\bibinfo  {journal} {Nanoscale}\ }\textbf {\bibinfo {volume} {7}},\ \bibinfo
  {pages} {17122} (\bibinfo {year} {2015})}\BibitemShut {NoStop}%
\bibitem [{\citenamefont {Bender}\ \emph
  {et~al.}(2018{\natexlab{a}})\citenamefont {Bender}, \citenamefont {Fock},
  \citenamefont {Frandsen}, \citenamefont {Hansen}, \citenamefont {Balceris},
  \citenamefont {Ludwig}, \citenamefont {Posth}, \citenamefont {Wetterskog},
  \citenamefont {Bogart}, \citenamefont {Southern}, \citenamefont {Szczerba},
  \citenamefont {Zeng}, \citenamefont {Witte}, \citenamefont {Gr\"uttner},
  \citenamefont {Westphal}, \citenamefont {Honecker}, \citenamefont
  {Gonz\'alez-Alonso}, \citenamefont {Fern\'andez~Barqu\'in},\ and\
  \citenamefont {Johansson}}]{bender2018jpcc}%
  \BibitemOpen
  \bibfield  {author} {\bibinfo {author} {\bibfnamefont {P.}~\bibnamefont
  {Bender}}, \bibinfo {author} {\bibfnamefont {J.}~\bibnamefont {Fock}},
  \bibinfo {author} {\bibfnamefont {C.}~\bibnamefont {Frandsen}}, \bibinfo
  {author} {\bibfnamefont {M.~F.}\ \bibnamefont {Hansen}}, \bibinfo {author}
  {\bibfnamefont {C.}~\bibnamefont {Balceris}}, \bibinfo {author}
  {\bibfnamefont {F.}~\bibnamefont {Ludwig}}, \bibinfo {author} {\bibfnamefont
  {O.}~\bibnamefont {Posth}}, \bibinfo {author} {\bibfnamefont
  {E.}~\bibnamefont {Wetterskog}}, \bibinfo {author} {\bibfnamefont {L.~K.}\
  \bibnamefont {Bogart}}, \bibinfo {author} {\bibfnamefont {P.}~\bibnamefont
  {Southern}}, \bibinfo {author} {\bibfnamefont {W.}~\bibnamefont {Szczerba}},
  \bibinfo {author} {\bibfnamefont {L.}~\bibnamefont {Zeng}}, \bibinfo {author}
  {\bibfnamefont {K.}~\bibnamefont {Witte}}, \bibinfo {author} {\bibfnamefont
  {C.}~\bibnamefont {Gr\"uttner}}, \bibinfo {author} {\bibfnamefont
  {F.}~\bibnamefont {Westphal}}, \bibinfo {author} {\bibfnamefont
  {D.}~\bibnamefont {Honecker}}, \bibinfo {author} {\bibfnamefont
  {D.}~\bibnamefont {Gonz\'alez-Alonso}}, \bibinfo {author} {\bibfnamefont
  {L.}~\bibnamefont {Fern\'andez~Barqu\'in}},\ and\ \bibinfo {author}
  {\bibfnamefont {C.}~\bibnamefont {Johansson}},\ }\href@noop {} {\bibfield
  {journal} {\bibinfo  {journal} {J. Phys. Chem. C}\ }\textbf {\bibinfo
  {volume} {122}},\ \bibinfo {pages} {3068} (\bibinfo {year}
  {2018}{\natexlab{a}})}\BibitemShut {NoStop}%
\bibitem [{\citenamefont {Bender}\ \emph
  {et~al.}(2018{\natexlab{b}})\citenamefont {Bender}, \citenamefont
  {Wetterskog}, \citenamefont {Honecker}, \citenamefont {Fock}, \citenamefont
  {Frandsen}, \citenamefont {Moerland}, \citenamefont {Bogart}, \citenamefont
  {Posth}, \citenamefont {Szczerba}, \citenamefont {Gavil{\'a}n}, \citenamefont
  {Costo}, \citenamefont {Fern\'andez-D\'iaz}, \citenamefont
  {Gonz\'alez-Alonso}, \citenamefont {Fern\'andez~Barqu\'in},\ and\
  \citenamefont {Johansson}}]{bender2018prb}%
  \BibitemOpen
  \bibfield  {author} {\bibinfo {author} {\bibfnamefont {P.}~\bibnamefont
  {Bender}}, \bibinfo {author} {\bibfnamefont {E.}~\bibnamefont {Wetterskog}},
  \bibinfo {author} {\bibfnamefont {D.}~\bibnamefont {Honecker}}, \bibinfo
  {author} {\bibfnamefont {J.}~\bibnamefont {Fock}}, \bibinfo {author}
  {\bibfnamefont {C.}~\bibnamefont {Frandsen}}, \bibinfo {author}
  {\bibfnamefont {C.}~\bibnamefont {Moerland}}, \bibinfo {author}
  {\bibfnamefont {L.~K.}\ \bibnamefont {Bogart}}, \bibinfo {author}
  {\bibfnamefont {O.}~\bibnamefont {Posth}}, \bibinfo {author} {\bibfnamefont
  {W.}~\bibnamefont {Szczerba}}, \bibinfo {author} {\bibfnamefont
  {H.}~\bibnamefont {Gavil{\'a}n}}, \bibinfo {author} {\bibfnamefont
  {R.}~\bibnamefont {Costo}}, \bibinfo {author} {\bibfnamefont {M.~T.}\
  \bibnamefont {Fern\'andez-D\'iaz}}, \bibinfo {author} {\bibfnamefont
  {D.}~\bibnamefont {Gonz\'alez-Alonso}}, \bibinfo {author} {\bibfnamefont
  {L.}~\bibnamefont {Fern\'andez~Barqu\'in}},\ and\ \bibinfo {author}
  {\bibfnamefont {C.}~\bibnamefont {Johansson}},\ }\href@noop {} {\bibfield
  {journal} {\bibinfo  {journal} {Phys. Rev. B}\ }\textbf {\bibinfo {volume}
  {98}},\ \bibinfo {pages} {224420} (\bibinfo {year}
  {2018}{\natexlab{b}})}\BibitemShut {NoStop}%
\bibitem [{\citenamefont {Oberdick}\ \emph {et~al.}(2018)\citenamefont
  {Oberdick}, \citenamefont {Abdelgawad}, \citenamefont {Moya}, \citenamefont
  {Mesbahi-Vasey}, \citenamefont {Kepaptsoglou}, \citenamefont {Lazarov},
  \citenamefont {Evans}, \citenamefont {Meilak}, \citenamefont {Skoropata},
  \citenamefont {van Lierop}, \citenamefont {Hunt-Isaak}, \citenamefont {Pan},
  \citenamefont {Ijiri}, \citenamefont {Krycka}, \citenamefont {Borchers},\
  and\ \citenamefont {Majetich}}]{oberdick2018}%
  \BibitemOpen
  \bibfield  {author} {\bibinfo {author} {\bibfnamefont {S.~D.}\ \bibnamefont
  {Oberdick}}, \bibinfo {author} {\bibfnamefont {A.}~\bibnamefont
  {Abdelgawad}}, \bibinfo {author} {\bibfnamefont {C.}~\bibnamefont {Moya}},
  \bibinfo {author} {\bibfnamefont {S.}~\bibnamefont {Mesbahi-Vasey}}, \bibinfo
  {author} {\bibfnamefont {D.}~\bibnamefont {Kepaptsoglou}}, \bibinfo {author}
  {\bibfnamefont {V.~K.}\ \bibnamefont {Lazarov}}, \bibinfo {author}
  {\bibfnamefont {R.~F.~L.}\ \bibnamefont {Evans}}, \bibinfo {author}
  {\bibfnamefont {D.}~\bibnamefont {Meilak}}, \bibinfo {author} {\bibfnamefont
  {E.}~\bibnamefont {Skoropata}}, \bibinfo {author} {\bibfnamefont
  {J.}~\bibnamefont {van Lierop}}, \bibinfo {author} {\bibfnamefont
  {I.}~\bibnamefont {Hunt-Isaak}}, \bibinfo {author} {\bibfnamefont
  {H.}~\bibnamefont {Pan}}, \bibinfo {author} {\bibfnamefont {Y.}~\bibnamefont
  {Ijiri}}, \bibinfo {author} {\bibfnamefont {K.~L.}\ \bibnamefont {Krycka}},
  \bibinfo {author} {\bibfnamefont {J.~A.}\ \bibnamefont {Borchers}},\ and\
  \bibinfo {author} {\bibfnamefont {S.~A.}\ \bibnamefont {Majetich}},\
  }\href@noop {} {\bibfield  {journal} {\bibinfo  {journal} {Sci. Rep.}\
  }\textbf {\bibinfo {volume} {8}},\ \bibinfo {pages} {3425} (\bibinfo {year}
  {2018})}\BibitemShut {NoStop}%
\bibitem [{\citenamefont {Ijiri}\ \emph {et~al.}(2019)\citenamefont {Ijiri},
  \citenamefont {Krycka}, \citenamefont {Hunt-Isaak}, \citenamefont {Pan},
  \citenamefont {Hsieh}, \citenamefont {Borchers}, \citenamefont {Rhyne},
  \citenamefont {Oberdick}, \citenamefont {Abdelgawad},\ and\ \citenamefont
  {Majetich}}]{krycka2019}%
  \BibitemOpen
  \bibfield  {author} {\bibinfo {author} {\bibfnamefont {Y.}~\bibnamefont
  {Ijiri}}, \bibinfo {author} {\bibfnamefont {K.~L.}\ \bibnamefont {Krycka}},
  \bibinfo {author} {\bibfnamefont {I.}~\bibnamefont {Hunt-Isaak}}, \bibinfo
  {author} {\bibfnamefont {H.}~\bibnamefont {Pan}}, \bibinfo {author}
  {\bibfnamefont {J.}~\bibnamefont {Hsieh}}, \bibinfo {author} {\bibfnamefont
  {J.~A.}\ \bibnamefont {Borchers}}, \bibinfo {author} {\bibfnamefont {J.~J.}\
  \bibnamefont {Rhyne}}, \bibinfo {author} {\bibfnamefont {S.~D.}\ \bibnamefont
  {Oberdick}}, \bibinfo {author} {\bibfnamefont {A.}~\bibnamefont
  {Abdelgawad}},\ and\ \bibinfo {author} {\bibfnamefont {S.~A.}\ \bibnamefont
  {Majetich}},\ }\href@noop {} {\bibfield  {journal} {\bibinfo  {journal}
  {Phys. Rev. B}\ }\textbf {\bibinfo {volume} {99}},\ \bibinfo {pages} {094421}
  (\bibinfo {year} {2019})}\BibitemShut {NoStop}%
\bibitem [{\citenamefont {Bender}\ \emph {et~al.}(2019)\citenamefont {Bender},
  \citenamefont {Honecker},\ and\ \citenamefont
  {Barqu\'{\i}n}}]{benderapl2019}%
  \BibitemOpen
  \bibfield  {author} {\bibinfo {author} {\bibfnamefont {P.}~\bibnamefont
  {Bender}}, \bibinfo {author} {\bibfnamefont {D.}~\bibnamefont {Honecker}},\
  and\ \bibinfo {author} {\bibfnamefont {L.~F.}\ \bibnamefont {Barqu\'{\i}n}},\
  }\href@noop {} {\bibfield  {journal} {\bibinfo  {journal} {Appl. Phys.
  Lett.}\ }\textbf {\bibinfo {volume} {115}},\ \bibinfo {pages} {132406}
  (\bibinfo {year} {2019})}\BibitemShut {NoStop}%
\bibitem [{\citenamefont {Bersweiler}\ \emph {et~al.}(2019)\citenamefont
  {Bersweiler}, \citenamefont {Bender}, \citenamefont {Vivas}, \citenamefont
  {Albino}, \citenamefont {Petrecca}, \citenamefont {M\"uhlbauer},
  \citenamefont {Erokhin}, \citenamefont {Berkov}, \citenamefont
  {Sangregorio},\ and\ \citenamefont {Michels}}]{bersweiler2019}%
  \BibitemOpen
  \bibfield  {author} {\bibinfo {author} {\bibfnamefont {M.}~\bibnamefont
  {Bersweiler}}, \bibinfo {author} {\bibfnamefont {P.}~\bibnamefont {Bender}},
  \bibinfo {author} {\bibfnamefont {L.~G.}\ \bibnamefont {Vivas}}, \bibinfo
  {author} {\bibfnamefont {M.}~\bibnamefont {Albino}}, \bibinfo {author}
  {\bibfnamefont {M.}~\bibnamefont {Petrecca}}, \bibinfo {author}
  {\bibfnamefont {S.}~\bibnamefont {M\"uhlbauer}}, \bibinfo {author}
  {\bibfnamefont {S.}~\bibnamefont {Erokhin}}, \bibinfo {author} {\bibfnamefont
  {D.}~\bibnamefont {Berkov}}, \bibinfo {author} {\bibfnamefont
  {C.}~\bibnamefont {Sangregorio}},\ and\ \bibinfo {author} {\bibfnamefont
  {A.}~\bibnamefont {Michels}},\ }\href@noop {} {\bibfield  {journal} {\bibinfo
   {journal} {Phys. Rev. B}\ }\textbf {\bibinfo {volume} {100}},\ \bibinfo
  {pages} {144434} (\bibinfo {year} {2019})}\BibitemShut {NoStop}%
\bibitem [{\citenamefont {Z\'akutn\'a}\ \emph {et~al.}(2020)\citenamefont
  {Z\'akutn\'a}, \citenamefont
  {Ni$\mathrm{\check{z}}$$\mathrm{\check{n}}$ansk\'y}, \citenamefont
  {Barnsley}, \citenamefont {Babcock}, \citenamefont {Salhi}, \citenamefont
  {Feoktystov}, \citenamefont {Honecker},\ and\ \citenamefont
  {Disch}}]{zakutna2020}%
  \BibitemOpen
  \bibfield  {author} {\bibinfo {author} {\bibfnamefont {D.}~\bibnamefont
  {Z\'akutn\'a}}, \bibinfo {author} {\bibfnamefont {D.}~\bibnamefont
  {Ni$\mathrm{\check{z}}$$\mathrm{\check{n}}$ansk\'y}}, \bibinfo {author}
  {\bibfnamefont {L.~C.}\ \bibnamefont {Barnsley}}, \bibinfo {author}
  {\bibfnamefont {E.}~\bibnamefont {Babcock}}, \bibinfo {author} {\bibfnamefont
  {Z.}~\bibnamefont {Salhi}}, \bibinfo {author} {\bibfnamefont
  {A.}~\bibnamefont {Feoktystov}}, \bibinfo {author} {\bibfnamefont
  {D.}~\bibnamefont {Honecker}},\ and\ \bibinfo {author} {\bibfnamefont
  {S.}~\bibnamefont {Disch}},\ }\href@noop {} {\bibfield  {journal} {\bibinfo
  {journal} {Phys. Rev. X}\ }\textbf {\bibinfo {volume} {10}},\ \bibinfo
  {pages} {031019} (\bibinfo {year} {2020})}\BibitemShut {NoStop}%
\bibitem [{\citenamefont {Vivas}\ \emph {et~al.}(2020)\citenamefont {Vivas},
  \citenamefont {Yanes}, \citenamefont {Berkov}, \citenamefont {Erokhin},
  \citenamefont {Bersweiler}, \citenamefont {Honecker}, \citenamefont
  {Bender},\ and\ \citenamefont {Michels}}]{laura2020}%
  \BibitemOpen
  \bibfield  {author} {\bibinfo {author} {\bibfnamefont {L.~G.}\ \bibnamefont
  {Vivas}}, \bibinfo {author} {\bibfnamefont {R.}~\bibnamefont {Yanes}},
  \bibinfo {author} {\bibfnamefont {D.}~\bibnamefont {Berkov}}, \bibinfo
  {author} {\bibfnamefont {S.}~\bibnamefont {Erokhin}}, \bibinfo {author}
  {\bibfnamefont {M.}~\bibnamefont {Bersweiler}}, \bibinfo {author}
  {\bibfnamefont {D.}~\bibnamefont {Honecker}}, \bibinfo {author}
  {\bibfnamefont {P.}~\bibnamefont {Bender}},\ and\ \bibinfo {author}
  {\bibfnamefont {A.}~\bibnamefont {Michels}},\ }\href@noop {} {\bibfield
  {journal} {\bibinfo  {journal} {Phys. Rev. Lett.}\ }\textbf {\bibinfo
  {volume} {125}},\ \bibinfo {pages} {117201} (\bibinfo {year}
  {2020})}\BibitemShut {NoStop}%
\bibitem [{\citenamefont {Ito}\ \emph {et~al.}(2007)\citenamefont {Ito},
  \citenamefont {Michels}, \citenamefont {Kohlbrecher}, \citenamefont
  {Garitaonandia}, \citenamefont {Suzuki},\ and\ \citenamefont
  {Cashion}}]{suzuki2007}%
  \BibitemOpen
  \bibfield  {author} {\bibinfo {author} {\bibfnamefont {N.}~\bibnamefont
  {Ito}}, \bibinfo {author} {\bibfnamefont {A.}~\bibnamefont {Michels}},
  \bibinfo {author} {\bibfnamefont {J.}~\bibnamefont {Kohlbrecher}}, \bibinfo
  {author} {\bibfnamefont {J.~S.}\ \bibnamefont {Garitaonandia}}, \bibinfo
  {author} {\bibfnamefont {K.}~\bibnamefont {Suzuki}},\ and\ \bibinfo {author}
  {\bibfnamefont {J.~D.}\ \bibnamefont {Cashion}},\ }\href@noop {} {\bibfield
  {journal} {\bibinfo  {journal} {J. Magn. Magn. Mater.}\ }\textbf {\bibinfo
  {volume} {316}},\ \bibinfo {pages} {458} (\bibinfo {year}
  {2007})}\BibitemShut {NoStop}%
\bibitem [{\citenamefont {Saranu}\ \emph {et~al.}(2008)\citenamefont {Saranu},
  \citenamefont {Grob}, \citenamefont {Weissm\"uller},\ and\ \citenamefont
  {Herr}}]{herr08pss}%
  \BibitemOpen
  \bibfield  {author} {\bibinfo {author} {\bibfnamefont {S.}~\bibnamefont
  {Saranu}}, \bibinfo {author} {\bibfnamefont {A.}~\bibnamefont {Grob}},
  \bibinfo {author} {\bibfnamefont {J.}~\bibnamefont {Weissm\"uller}},\ and\
  \bibinfo {author} {\bibfnamefont {U.}~\bibnamefont {Herr}},\ }\href@noop {}
  {\bibfield  {journal} {\bibinfo  {journal} {Phys. Status Solidi A}\ }\textbf
  {\bibinfo {volume} {205}},\ \bibinfo {pages} {1774} (\bibinfo {year}
  {2008})}\BibitemShut {NoStop}%
\bibitem [{\citenamefont {Mettus}\ \emph {et~al.}(2017)\citenamefont {Mettus},
  \citenamefont {Deckarm}, \citenamefont {Leibner}, \citenamefont {Birringer},
  \citenamefont {Stolpe}, \citenamefont {Busch}, \citenamefont {Honecker},
  \citenamefont {Kohlbrecher}, \citenamefont {Hautle}, \citenamefont {Niketic},
  \citenamefont {Fern\'andez}, \citenamefont {Barqu\'{\i}n},\ and\
  \citenamefont {Michels}}]{mettusprm2017}%
  \BibitemOpen
  \bibfield  {author} {\bibinfo {author} {\bibfnamefont {D.}~\bibnamefont
  {Mettus}}, \bibinfo {author} {\bibfnamefont {M.}~\bibnamefont {Deckarm}},
  \bibinfo {author} {\bibfnamefont {A.}~\bibnamefont {Leibner}}, \bibinfo
  {author} {\bibfnamefont {R.}~\bibnamefont {Birringer}}, \bibinfo {author}
  {\bibfnamefont {M.}~\bibnamefont {Stolpe}}, \bibinfo {author} {\bibfnamefont
  {R.}~\bibnamefont {Busch}}, \bibinfo {author} {\bibfnamefont
  {D.}~\bibnamefont {Honecker}}, \bibinfo {author} {\bibfnamefont
  {J.}~\bibnamefont {Kohlbrecher}}, \bibinfo {author} {\bibfnamefont
  {P.}~\bibnamefont {Hautle}}, \bibinfo {author} {\bibfnamefont
  {N.}~\bibnamefont {Niketic}}, \bibinfo {author} {\bibfnamefont {J.~R.}\
  \bibnamefont {Fern\'andez}}, \bibinfo {author} {\bibfnamefont {L.~F.}\
  \bibnamefont {Barqu\'{\i}n}},\ and\ \bibinfo {author} {\bibfnamefont
  {A.}~\bibnamefont {Michels}},\ }\href@noop {} {\bibfield  {journal} {\bibinfo
   {journal} {Phys. Rev. Materials}\ }\textbf {\bibinfo {volume} {1}},\
  \bibinfo {pages} {074403} (\bibinfo {year} {2017})}\BibitemShut {NoStop}%
\bibitem [{\citenamefont {Mirebeau}\ \emph {et~al.}(2018)\citenamefont
  {Mirebeau}, \citenamefont {Martin}, \citenamefont {Deutsch}, \citenamefont
  {Bannenberg}, \citenamefont {Pappas}, \citenamefont {Chaboussant},
  \citenamefont {Cubitt}, \citenamefont {Decorse},\ and\ \citenamefont
  {Leonov}}]{mirebeau2018}%
  \BibitemOpen
  \bibfield  {author} {\bibinfo {author} {\bibfnamefont {I.}~\bibnamefont
  {Mirebeau}}, \bibinfo {author} {\bibfnamefont {N.}~\bibnamefont {Martin}},
  \bibinfo {author} {\bibfnamefont {M.}~\bibnamefont {Deutsch}}, \bibinfo
  {author} {\bibfnamefont {L.~J.}\ \bibnamefont {Bannenberg}}, \bibinfo
  {author} {\bibfnamefont {C.}~\bibnamefont {Pappas}}, \bibinfo {author}
  {\bibfnamefont {G.}~\bibnamefont {Chaboussant}}, \bibinfo {author}
  {\bibfnamefont {R.}~\bibnamefont {Cubitt}}, \bibinfo {author} {\bibfnamefont
  {C.}~\bibnamefont {Decorse}},\ and\ \bibinfo {author} {\bibfnamefont {A.~O.}\
  \bibnamefont {Leonov}},\ }\href@noop {} {\bibfield  {journal} {\bibinfo
  {journal} {Phys. Rev. B}\ }\textbf {\bibinfo {volume} {98}},\ \bibinfo
  {pages} {014420} (\bibinfo {year} {2018})}\BibitemShut {NoStop}%
\bibitem [{\citenamefont {Schroeder}\ \emph {et~al.}(2020)\citenamefont
  {Schroeder}, \citenamefont {Bhattarai}, \citenamefont {Gebretsadik},
  \citenamefont {Adawi}, \citenamefont {Lussier},\ and\ \citenamefont
  {Krycka}}]{schroeder2020}%
  \BibitemOpen
  \bibfield  {author} {\bibinfo {author} {\bibfnamefont {A.}~\bibnamefont
  {Schroeder}}, \bibinfo {author} {\bibfnamefont {S.}~\bibnamefont
  {Bhattarai}}, \bibinfo {author} {\bibfnamefont {A.}~\bibnamefont
  {Gebretsadik}}, \bibinfo {author} {\bibfnamefont {H.}~\bibnamefont {Adawi}},
  \bibinfo {author} {\bibfnamefont {J.-G.}\ \bibnamefont {Lussier}},\ and\
  \bibinfo {author} {\bibfnamefont {K.~L.}\ \bibnamefont {Krycka}},\
  }\href@noop {} {\bibfield  {journal} {\bibinfo  {journal} {AIP Advances}\
  }\textbf {\bibinfo {volume} {10}},\ \bibinfo {pages} {015036} (\bibinfo
  {year} {2020})}\BibitemShut {NoStop}%
\bibitem [{\citenamefont {Bersweiler}\ \emph {et~al.}(2020)\citenamefont
  {Bersweiler}, \citenamefont {Bender}, \citenamefont {Peral}, \citenamefont
  {Eichenberger}, \citenamefont {Hehn}, \citenamefont {Polewczyk},
  \citenamefont {M\"uhlbauer},\ and\ \citenamefont {Michels}}]{bersweiler2020}%
  \BibitemOpen
  \bibfield  {author} {\bibinfo {author} {\bibfnamefont {M.}~\bibnamefont
  {Bersweiler}}, \bibinfo {author} {\bibfnamefont {P.}~\bibnamefont {Bender}},
  \bibinfo {author} {\bibfnamefont {I.}~\bibnamefont {Peral}}, \bibinfo
  {author} {\bibfnamefont {L.}~\bibnamefont {Eichenberger}}, \bibinfo {author}
  {\bibfnamefont {M.}~\bibnamefont {Hehn}}, \bibinfo {author} {\bibfnamefont
  {V.}~\bibnamefont {Polewczyk}}, \bibinfo {author} {\bibfnamefont
  {S.}~\bibnamefont {M\"uhlbauer}},\ and\ \bibinfo {author} {\bibfnamefont
  {A.}~\bibnamefont {Michels}},\ }\href@noop {} {\bibfield  {journal} {\bibinfo
   {journal} {J. Phys. D: Appl. Phys.}\ }\textbf {\bibinfo {volume} {53}},\
  \bibinfo {pages} {335302} (\bibinfo {year} {2020})}\BibitemShut {NoStop}%
\bibitem [{\citenamefont {Oba}\ \emph {et~al.}(2020)\citenamefont {Oba},
  \citenamefont {Adachi}, \citenamefont {Todaka}, \citenamefont {Gilbert},\
  and\ \citenamefont {Mamiya}}]{oba2020}%
  \BibitemOpen
  \bibfield  {author} {\bibinfo {author} {\bibfnamefont {Y.}~\bibnamefont
  {Oba}}, \bibinfo {author} {\bibfnamefont {N.}~\bibnamefont {Adachi}},
  \bibinfo {author} {\bibfnamefont {Y.}~\bibnamefont {Todaka}}, \bibinfo
  {author} {\bibfnamefont {E.~P.}\ \bibnamefont {Gilbert}},\ and\ \bibinfo
  {author} {\bibfnamefont {H.}~\bibnamefont {Mamiya}},\ }\href@noop {}
  {\bibfield  {journal} {\bibinfo  {journal} {Phys. Rev. Research}\ }\textbf
  {\bibinfo {volume} {2}},\ \bibinfo {pages} {033473} (\bibinfo {year}
  {2020})}\BibitemShut {NoStop}%
\bibitem [{\citenamefont {van~den Brandt}\ \emph {et~al.}(2006)\citenamefont
  {van~den Brandt}, \citenamefont {Gl\"attli}, \citenamefont {Grillo},
  \citenamefont {Hautle}, \citenamefont {Jouve}, \citenamefont {Kohlbrecher},
  \citenamefont {Konter}, \citenamefont {Leymarie}, \citenamefont {Mango},
  \citenamefont {May}, \citenamefont {Michels}, \citenamefont {Stuhrmann},\
  and\ \citenamefont {Zimmer}}]{michels06a}%
  \BibitemOpen
  \bibfield  {author} {\bibinfo {author} {\bibfnamefont {B.}~\bibnamefont
  {van~den Brandt}}, \bibinfo {author} {\bibfnamefont {H.}~\bibnamefont
  {Gl\"attli}}, \bibinfo {author} {\bibfnamefont {I.}~\bibnamefont {Grillo}},
  \bibinfo {author} {\bibfnamefont {P.}~\bibnamefont {Hautle}}, \bibinfo
  {author} {\bibfnamefont {H.}~\bibnamefont {Jouve}}, \bibinfo {author}
  {\bibfnamefont {J.}~\bibnamefont {Kohlbrecher}}, \bibinfo {author}
  {\bibfnamefont {J.~A.}\ \bibnamefont {Konter}}, \bibinfo {author}
  {\bibfnamefont {E.}~\bibnamefont {Leymarie}}, \bibinfo {author}
  {\bibfnamefont {S.}~\bibnamefont {Mango}}, \bibinfo {author} {\bibfnamefont
  {R.~P.}\ \bibnamefont {May}}, \bibinfo {author} {\bibfnamefont
  {A.}~\bibnamefont {Michels}}, \bibinfo {author} {\bibfnamefont {H.~B.}\
  \bibnamefont {Stuhrmann}},\ and\ \bibinfo {author} {\bibfnamefont
  {O.}~\bibnamefont {Zimmer}},\ }\href@noop {} {\bibfield  {journal} {\bibinfo
  {journal} {Eur. Phys. J. B}\ }\textbf {\bibinfo {volume} {49}},\ \bibinfo
  {pages} {157} (\bibinfo {year} {2006})}\BibitemShut {NoStop}%
\bibitem [{\citenamefont {Aswal}\ \emph {et~al.}(2008)\citenamefont {Aswal},
  \citenamefont {van~den Brandt}, \citenamefont {Hautle}, \citenamefont
  {Kohlbrecher}, \citenamefont {Konter}, \citenamefont {Michels}, \citenamefont
  {Piegsa}, \citenamefont {Stahn}, \citenamefont {{S. Van Petegem}},\ and\
  \citenamefont {Zimmer}}]{aswal08nim}%
  \BibitemOpen
  \bibfield  {author} {\bibinfo {author} {\bibfnamefont {V.~K.}\ \bibnamefont
  {Aswal}}, \bibinfo {author} {\bibfnamefont {B.}~\bibnamefont {van~den
  Brandt}}, \bibinfo {author} {\bibfnamefont {P.}~\bibnamefont {Hautle}},
  \bibinfo {author} {\bibfnamefont {J.}~\bibnamefont {Kohlbrecher}}, \bibinfo
  {author} {\bibfnamefont {J.~A.}\ \bibnamefont {Konter}}, \bibinfo {author}
  {\bibfnamefont {A.}~\bibnamefont {Michels}}, \bibinfo {author} {\bibfnamefont
  {F.~M.}\ \bibnamefont {Piegsa}}, \bibinfo {author} {\bibfnamefont
  {J.}~\bibnamefont {Stahn}}, \bibinfo {author} {\bibnamefont {{S. Van
  Petegem}}},\ and\ \bibinfo {author} {\bibfnamefont {O.}~\bibnamefont
  {Zimmer}},\ }\href@noop {} {\bibfield  {journal} {\bibinfo  {journal} {Nucl.
  Instrum. Methods Phys. Res. A}\ }\textbf {\bibinfo {volume} {586}},\ \bibinfo
  {pages} {86} (\bibinfo {year} {2008})}\BibitemShut {NoStop}%
\bibitem [{\citenamefont {Noda}\ \emph {et~al.}(2016)\citenamefont {Noda},
  \citenamefont {Koizumi}, \citenamefont {Masui}, \citenamefont {Mashita},
  \citenamefont {Kishimoto}, \citenamefont {Yamaguchi}, \citenamefont {Kumada},
  \citenamefont {Takata}, \citenamefont {Ohishi},\ and\ \citenamefont
  {Suzuki}}]{noda2016}%
  \BibitemOpen
  \bibfield  {author} {\bibinfo {author} {\bibfnamefont {Y.}~\bibnamefont
  {Noda}}, \bibinfo {author} {\bibfnamefont {S.}~\bibnamefont {Koizumi}},
  \bibinfo {author} {\bibfnamefont {T.}~\bibnamefont {Masui}}, \bibinfo
  {author} {\bibfnamefont {R.}~\bibnamefont {Mashita}}, \bibinfo {author}
  {\bibfnamefont {H.}~\bibnamefont {Kishimoto}}, \bibinfo {author}
  {\bibfnamefont {D.}~\bibnamefont {Yamaguchi}}, \bibinfo {author}
  {\bibfnamefont {T.}~\bibnamefont {Kumada}}, \bibinfo {author} {\bibfnamefont
  {S.-i.}\ \bibnamefont {Takata}}, \bibinfo {author} {\bibfnamefont
  {K.}~\bibnamefont {Ohishi}},\ and\ \bibinfo {author} {\bibfnamefont
  {J.}~\bibnamefont {Suzuki}},\ }\href@noop {} {\bibfield  {journal} {\bibinfo
  {journal} {J. Appl. Cryst.}\ }\textbf {\bibinfo {volume} {49}},\ \bibinfo
  {pages} {2036} (\bibinfo {year} {2016})}\BibitemShut {NoStop}%
\bibitem [{\citenamefont {Bischof}\ \emph {et~al.}(2007)\citenamefont
  {Bischof}, \citenamefont {Staron}, \citenamefont {Michels}, \citenamefont
  {Granitzer}, \citenamefont {Rumpf}, \citenamefont {Leitner}, \citenamefont
  {Scheu},\ and\ \citenamefont {Clemens}}]{bischof07}%
  \BibitemOpen
  \bibfield  {author} {\bibinfo {author} {\bibfnamefont {M.}~\bibnamefont
  {Bischof}}, \bibinfo {author} {\bibfnamefont {P.}~\bibnamefont {Staron}},
  \bibinfo {author} {\bibfnamefont {A.}~\bibnamefont {Michels}}, \bibinfo
  {author} {\bibfnamefont {P.}~\bibnamefont {Granitzer}}, \bibinfo {author}
  {\bibfnamefont {K.}~\bibnamefont {Rumpf}}, \bibinfo {author} {\bibfnamefont
  {H.}~\bibnamefont {Leitner}}, \bibinfo {author} {\bibfnamefont
  {C.}~\bibnamefont {Scheu}},\ and\ \bibinfo {author} {\bibfnamefont
  {H.}~\bibnamefont {Clemens}},\ }\href@noop {} {\bibfield  {journal} {\bibinfo
   {journal} {Acta Mater.}\ }\textbf {\bibinfo {volume} {55}},\ \bibinfo
  {pages} {2637} (\bibinfo {year} {2007})}\BibitemShut {NoStop}%
\bibitem [{\citenamefont {Bergner}\ \emph {et~al.}(2013)\citenamefont
  {Bergner}, \citenamefont {Pareige}, \citenamefont {Kuksenko}, \citenamefont
  {Malerba}, \citenamefont {Pareige}, \citenamefont {Ulbricht},\ and\
  \citenamefont {Wagner}}]{bergner2013}%
  \BibitemOpen
  \bibfield  {author} {\bibinfo {author} {\bibfnamefont {F.}~\bibnamefont
  {Bergner}}, \bibinfo {author} {\bibfnamefont {C.}~\bibnamefont {Pareige}},
  \bibinfo {author} {\bibfnamefont {V.}~\bibnamefont {Kuksenko}}, \bibinfo
  {author} {\bibfnamefont {L.}~\bibnamefont {Malerba}}, \bibinfo {author}
  {\bibfnamefont {P.}~\bibnamefont {Pareige}}, \bibinfo {author} {\bibfnamefont
  {A.}~\bibnamefont {Ulbricht}},\ and\ \bibinfo {author} {\bibfnamefont
  {A.}~\bibnamefont {Wagner}},\ }\href@noop {} {\bibfield  {journal} {\bibinfo
  {journal} {J. Nucl. Mater.}\ }\textbf {\bibinfo {volume} {442}},\ \bibinfo
  {pages} {463} (\bibinfo {year} {2013})}\BibitemShut {NoStop}%
\bibitem [{\citenamefont {Pareja}\ \emph {et~al.}(2015)\citenamefont {Pareja},
  \citenamefont {Parente}, \citenamefont {Mu{\~{n}}oz}, \citenamefont
  {Radulescu},\ and\ \citenamefont {de~Castro}}]{Pareja:15}%
  \BibitemOpen
  \bibfield  {author} {\bibinfo {author} {\bibfnamefont {R.}~\bibnamefont
  {Pareja}}, \bibinfo {author} {\bibfnamefont {P.}~\bibnamefont {Parente}},
  \bibinfo {author} {\bibfnamefont {A.}~\bibnamefont {Mu{\~{n}}oz}}, \bibinfo
  {author} {\bibfnamefont {A.}~\bibnamefont {Radulescu}},\ and\ \bibinfo
  {author} {\bibfnamefont {V.}~\bibnamefont {de~Castro}},\ }\href@noop {}
  {\bibfield  {journal} {\bibinfo  {journal} {Philos. Mag.}\ }\textbf {\bibinfo
  {volume} {95}},\ \bibinfo {pages} {2450} (\bibinfo {year}
  {2015})}\BibitemShut {NoStop}%
\bibitem [{\citenamefont {Oba}\ \emph {et~al.}(2016)\citenamefont {Oba},
  \citenamefont {Morooka}, \citenamefont {Ohishi}, \citenamefont {Sato},
  \citenamefont {Inoue}, \citenamefont {Adachi}, \citenamefont {Suzuki},
  \citenamefont {Tsuchiyama}, \citenamefont {Gilbert},\ and\ \citenamefont
  {Sugiyama}}]{gilbert2016}%
  \BibitemOpen
  \bibfield  {author} {\bibinfo {author} {\bibfnamefont {Y.}~\bibnamefont
  {Oba}}, \bibinfo {author} {\bibfnamefont {S.}~\bibnamefont {Morooka}},
  \bibinfo {author} {\bibfnamefont {K.}~\bibnamefont {Ohishi}}, \bibinfo
  {author} {\bibfnamefont {N.}~\bibnamefont {Sato}}, \bibinfo {author}
  {\bibfnamefont {R.}~\bibnamefont {Inoue}}, \bibinfo {author} {\bibfnamefont
  {N.}~\bibnamefont {Adachi}}, \bibinfo {author} {\bibfnamefont
  {J.}~\bibnamefont {Suzuki}}, \bibinfo {author} {\bibfnamefont
  {T.}~\bibnamefont {Tsuchiyama}}, \bibinfo {author} {\bibfnamefont {E.~P.}\
  \bibnamefont {Gilbert}},\ and\ \bibinfo {author} {\bibfnamefont
  {M.}~\bibnamefont {Sugiyama}},\ }\href@noop {} {\bibfield  {journal}
  {\bibinfo  {journal} {J. Appl. Cryst.}\ }\textbf {\bibinfo {volume} {49}},\
  \bibinfo {pages} {1659} (\bibinfo {year} {2016})}\BibitemShut {NoStop}%
\bibitem [{\citenamefont {Shu}\ \emph {et~al.}(2018)\citenamefont {Shu},
  \citenamefont {Wirth}, \citenamefont {Wells}, \citenamefont {Morgan},\ and\
  \citenamefont {Odette}}]{shu2018}%
  \BibitemOpen
  \bibfield  {author} {\bibinfo {author} {\bibfnamefont {S.}~\bibnamefont
  {Shu}}, \bibinfo {author} {\bibfnamefont {B.~D.}\ \bibnamefont {Wirth}},
  \bibinfo {author} {\bibfnamefont {P.~B.}\ \bibnamefont {Wells}}, \bibinfo
  {author} {\bibfnamefont {D.~D.}\ \bibnamefont {Morgan}},\ and\ \bibinfo
  {author} {\bibfnamefont {G.~R.}\ \bibnamefont {Odette}},\ }\href@noop {}
  {\bibfield  {journal} {\bibinfo  {journal} {Acta Mater.}\ }\textbf {\bibinfo
  {volume} {146}},\ \bibinfo {pages} {237} (\bibinfo {year}
  {2018})}\BibitemShut {NoStop}%
\bibitem [{\citenamefont {Bhatti}\ \emph {et~al.}(2012)\citenamefont {Bhatti},
  \citenamefont {El-Khatib}, \citenamefont {Srivastava}, \citenamefont
  {James},\ and\ \citenamefont {Leighton}}]{bhatti2012}%
  \BibitemOpen
  \bibfield  {author} {\bibinfo {author} {\bibfnamefont {K.~P.}\ \bibnamefont
  {Bhatti}}, \bibinfo {author} {\bibfnamefont {S.}~\bibnamefont {El-Khatib}},
  \bibinfo {author} {\bibfnamefont {V.}~\bibnamefont {Srivastava}}, \bibinfo
  {author} {\bibfnamefont {R.~D.}\ \bibnamefont {James}},\ and\ \bibinfo
  {author} {\bibfnamefont {C.}~\bibnamefont {Leighton}},\ }\href@noop {}
  {\bibfield  {journal} {\bibinfo  {journal} {Phys. Rev. B}\ }\textbf {\bibinfo
  {volume} {85}},\ \bibinfo {pages} {134450} (\bibinfo {year}
  {2012})}\BibitemShut {NoStop}%
\bibitem [{\citenamefont {Runov}\ \emph {et~al.}(2006)\citenamefont {Runov},
  \citenamefont {{Yu. P. Chernenkov}}, \citenamefont {Runova}, \citenamefont
  {Gavrilyuk}, \citenamefont {Glavatska}, \citenamefont {Goukasov},
  \citenamefont {Koledov}, \citenamefont {Shavrov},\ and\ \citenamefont
  {Khova{\u{\i}}lo}}]{runov2006}%
  \BibitemOpen
  \bibfield  {author} {\bibinfo {author} {\bibfnamefont {V.~V.}\ \bibnamefont
  {Runov}}, \bibinfo {author} {\bibnamefont {{Yu. P. Chernenkov}}}, \bibinfo
  {author} {\bibfnamefont {M.~K.}\ \bibnamefont {Runova}}, \bibinfo {author}
  {\bibfnamefont {V.~G.}\ \bibnamefont {Gavrilyuk}}, \bibinfo {author}
  {\bibfnamefont {N.~I.}\ \bibnamefont {Glavatska}}, \bibinfo {author}
  {\bibfnamefont {A.~G.}\ \bibnamefont {Goukasov}}, \bibinfo {author}
  {\bibfnamefont {V.~V.}\ \bibnamefont {Koledov}}, \bibinfo {author}
  {\bibfnamefont {V.~G.}\ \bibnamefont {Shavrov}},\ and\ \bibinfo {author}
  {\bibfnamefont {V.~V.}\ \bibnamefont {Khova{\u{\i}}lo}},\ }\href@noop {}
  {\bibfield  {journal} {\bibinfo  {journal} {J. Exp. Theo. Phys.}\ }\textbf
  {\bibinfo {volume} {102}},\ \bibinfo {pages} {102} (\bibinfo {year}
  {2006})}\BibitemShut {NoStop}%
\bibitem [{\citenamefont {Benacchio}\ \emph {et~al.}(2019)\citenamefont
  {Benacchio}, \citenamefont {Titov}, \citenamefont {Malyeyev}, \citenamefont
  {Peral}, \citenamefont {Bersweiler}, \citenamefont {Bender}, \citenamefont
  {Mettus}, \citenamefont {Honecker}, \citenamefont {Gilbert}, \citenamefont
  {Coduri}, \citenamefont {Heinemann}, \citenamefont {M\"uhlbauer},
  \citenamefont {{\c{C}}ak{\i}r}, \citenamefont {Acet},\ and\ \citenamefont
  {Michels}}]{michelsheusler2019}%
  \BibitemOpen
  \bibfield  {author} {\bibinfo {author} {\bibfnamefont {G.}~\bibnamefont
  {Benacchio}}, \bibinfo {author} {\bibfnamefont {I.}~\bibnamefont {Titov}},
  \bibinfo {author} {\bibfnamefont {A.}~\bibnamefont {Malyeyev}}, \bibinfo
  {author} {\bibfnamefont {I.}~\bibnamefont {Peral}}, \bibinfo {author}
  {\bibfnamefont {M.}~\bibnamefont {Bersweiler}}, \bibinfo {author}
  {\bibfnamefont {P.}~\bibnamefont {Bender}}, \bibinfo {author} {\bibfnamefont
  {D.}~\bibnamefont {Mettus}}, \bibinfo {author} {\bibfnamefont
  {D.}~\bibnamefont {Honecker}}, \bibinfo {author} {\bibfnamefont {E.~P.}\
  \bibnamefont {Gilbert}}, \bibinfo {author} {\bibfnamefont {M.}~\bibnamefont
  {Coduri}}, \bibinfo {author} {\bibfnamefont {A.}~\bibnamefont {Heinemann}},
  \bibinfo {author} {\bibfnamefont {S.}~\bibnamefont {M\"uhlbauer}}, \bibinfo
  {author} {\bibfnamefont {A.}~\bibnamefont {{\c{C}}ak{\i}r}}, \bibinfo
  {author} {\bibfnamefont {M.}~\bibnamefont {Acet}},\ and\ \bibinfo {author}
  {\bibfnamefont {A.}~\bibnamefont {Michels}},\ }\href@noop {} {\bibfield
  {journal} {\bibinfo  {journal} {Phys. Rev. B}\ }\textbf {\bibinfo {volume}
  {99}},\ \bibinfo {pages} {184422} (\bibinfo {year} {2019})}\BibitemShut
  {NoStop}%
\bibitem [{\citenamefont {El-Khatib}\ \emph {et~al.}(2019)\citenamefont
  {El-Khatib}, \citenamefont {Bhatti}, \citenamefont {Srivastava},
  \citenamefont {James},\ and\ \citenamefont {Leighton}}]{leighton2019}%
  \BibitemOpen
  \bibfield  {author} {\bibinfo {author} {\bibfnamefont {S.}~\bibnamefont
  {El-Khatib}}, \bibinfo {author} {\bibfnamefont {K.~P.}\ \bibnamefont
  {Bhatti}}, \bibinfo {author} {\bibfnamefont {V.}~\bibnamefont {Srivastava}},
  \bibinfo {author} {\bibfnamefont {R.~D.}\ \bibnamefont {James}},\ and\
  \bibinfo {author} {\bibfnamefont {C.}~\bibnamefont {Leighton}},\ }\href@noop
  {} {\bibfield  {journal} {\bibinfo  {journal} {Phys. Rev. Materials}\
  }\textbf {\bibinfo {volume} {3}},\ \bibinfo {pages} {104413} (\bibinfo {year}
  {2019})}\BibitemShut {NoStop}%
\bibitem [{\citenamefont {Sarkar}\ \emph {et~al.}(2020)\citenamefont {Sarkar},
  \citenamefont {Ahlawat}, \citenamefont {Kaushik}, \citenamefont {Babu},
  \citenamefont {Sen}, \citenamefont {Honecker},\ and\ \citenamefont
  {Biswas}}]{sarkar2020}%
  \BibitemOpen
  \bibfield  {author} {\bibinfo {author} {\bibfnamefont {S.~K.}\ \bibnamefont
  {Sarkar}}, \bibinfo {author} {\bibfnamefont {S.}~\bibnamefont {Ahlawat}},
  \bibinfo {author} {\bibfnamefont {S.~D.}\ \bibnamefont {Kaushik}}, \bibinfo
  {author} {\bibfnamefont {P.~D.}\ \bibnamefont {Babu}}, \bibinfo {author}
  {\bibfnamefont {D.}~\bibnamefont {Sen}}, \bibinfo {author} {\bibfnamefont
  {D.}~\bibnamefont {Honecker}},\ and\ \bibinfo {author} {\bibfnamefont
  {A.}~\bibnamefont {Biswas}},\ }\href@noop {} {\bibfield  {journal} {\bibinfo
  {journal} {J. Phys.: Condens. Matter}\ }\textbf {\bibinfo {volume} {32}},\
  \bibinfo {pages} {115801} (\bibinfo {year} {2020})}\BibitemShut {NoStop}%
\bibitem [{\citenamefont {M{\"{u}}hlbauer}\ \emph {et~al.}(2019)\citenamefont
  {M{\"{u}}hlbauer}, \citenamefont {Honecker}, \citenamefont {P{\'{e}}rigo},
  \citenamefont {Bergner}, \citenamefont {Disch}, \citenamefont {Heinemann},
  \citenamefont {Erokhin}, \citenamefont {Berkov}, \citenamefont {Leighton},
  \citenamefont {Eskildsen},\ and\ \citenamefont {Michels}}]{Muhlbauer2019}%
  \BibitemOpen
  \bibfield  {author} {\bibinfo {author} {\bibfnamefont {S.}~\bibnamefont
  {M{\"{u}}hlbauer}}, \bibinfo {author} {\bibfnamefont {D.}~\bibnamefont
  {Honecker}}, \bibinfo {author} {\bibfnamefont {{\'{E}}.~A.}\ \bibnamefont
  {P{\'{e}}rigo}}, \bibinfo {author} {\bibfnamefont {F.}~\bibnamefont
  {Bergner}}, \bibinfo {author} {\bibfnamefont {S.}~\bibnamefont {Disch}},
  \bibinfo {author} {\bibfnamefont {A.}~\bibnamefont {Heinemann}}, \bibinfo
  {author} {\bibfnamefont {S.}~\bibnamefont {Erokhin}}, \bibinfo {author}
  {\bibfnamefont {D.}~\bibnamefont {Berkov}}, \bibinfo {author} {\bibfnamefont
  {C.}~\bibnamefont {Leighton}}, \bibinfo {author} {\bibfnamefont {M.~R.}\
  \bibnamefont {Eskildsen}},\ and\ \bibinfo {author} {\bibfnamefont
  {A.}~\bibnamefont {Michels}},\ }\href
  {https://doi.org/10.1103/RevModPhys.91.015004} {\bibfield  {journal}
  {\bibinfo  {journal} {Rev. Mod. Phys.}\ }\textbf {\bibinfo {volume} {91}},\
  \bibinfo {pages} {015004} (\bibinfo {year} {2019})}\BibitemShut {NoStop}%
\bibitem [{sup()}]{suppmaterial}%
  \BibitemOpen
  \href@noop {} {}\bibinfo {note} {See the Supplemental Material [URL] for
  further neutron data.}\BibitemShut {Stop}%
\bibitem [{\citenamefont {Chen}\ \emph {et~al.}(2016)\citenamefont {Chen},
  \citenamefont {Sawatzki}, \citenamefont {Ener}, \citenamefont {Sepehri-Amin},
  \citenamefont {Leineweber}, \citenamefont {Gregori}, \citenamefont {Qu},
  \citenamefont {Muralidhar}, \citenamefont {Ohkubo}, \citenamefont {Hono},
  \citenamefont {Gutfleisch}, \citenamefont {Kronm{\"{u}}ller}, \citenamefont
  {Sch{\"{u}}tz},\ and\ \citenamefont {Goering}}]{Chen2016}%
  \BibitemOpen
  \bibfield  {author} {\bibinfo {author} {\bibfnamefont {Y.-C.}\ \bibnamefont
  {Chen}}, \bibinfo {author} {\bibfnamefont {S.}~\bibnamefont {Sawatzki}},
  \bibinfo {author} {\bibfnamefont {S.}~\bibnamefont {Ener}}, \bibinfo {author}
  {\bibfnamefont {H.}~\bibnamefont {Sepehri-Amin}}, \bibinfo {author}
  {\bibfnamefont {A.}~\bibnamefont {Leineweber}}, \bibinfo {author}
  {\bibfnamefont {G.}~\bibnamefont {Gregori}}, \bibinfo {author} {\bibfnamefont
  {F.}~\bibnamefont {Qu}}, \bibinfo {author} {\bibfnamefont {S.}~\bibnamefont
  {Muralidhar}}, \bibinfo {author} {\bibfnamefont {T.}~\bibnamefont {Ohkubo}},
  \bibinfo {author} {\bibfnamefont {K.}~\bibnamefont {Hono}}, \bibinfo {author}
  {\bibfnamefont {O.}~\bibnamefont {Gutfleisch}}, \bibinfo {author}
  {\bibfnamefont {H.}~\bibnamefont {Kronm{\"{u}}ller}}, \bibinfo {author}
  {\bibfnamefont {G.}~\bibnamefont {Sch{\"{u}}tz}},\ and\ \bibinfo {author}
  {\bibfnamefont {E.}~\bibnamefont {Goering}},\ }\href
  {https://doi.org/10.1063/1.4971759} {\bibfield  {journal} {\bibinfo
  {journal} {AIP Advances}\ }\textbf {\bibinfo {volume} {6}},\ \bibinfo {pages}
  {125301} (\bibinfo {year} {2016})}\BibitemShut {NoStop}%
\bibitem [{\citenamefont {Muralidhar}\ \emph {et~al.}(2017)\citenamefont
  {Muralidhar}, \citenamefont {Gr{\"{a}}fe}, \citenamefont {Chen},
  \citenamefont {Etter}, \citenamefont {Gregori}, \citenamefont {Ener},
  \citenamefont {Sawatzki}, \citenamefont {Hono}, \citenamefont {Gutfleisch},
  \citenamefont {Kronm{\"{u}}ller}, \citenamefont {Sch{\"{u}}tz},\ and\
  \citenamefont {Goering}}]{Muralidhar2017}%
  \BibitemOpen
  \bibfield  {author} {\bibinfo {author} {\bibfnamefont {S.}~\bibnamefont
  {Muralidhar}}, \bibinfo {author} {\bibfnamefont {J.}~\bibnamefont
  {Gr{\"{a}}fe}}, \bibinfo {author} {\bibfnamefont {Y.~C.}\ \bibnamefont
  {Chen}}, \bibinfo {author} {\bibfnamefont {M.}~\bibnamefont {Etter}},
  \bibinfo {author} {\bibfnamefont {G.}~\bibnamefont {Gregori}}, \bibinfo
  {author} {\bibfnamefont {S.}~\bibnamefont {Ener}}, \bibinfo {author}
  {\bibfnamefont {S.}~\bibnamefont {Sawatzki}}, \bibinfo {author}
  {\bibfnamefont {K.}~\bibnamefont {Hono}}, \bibinfo {author} {\bibfnamefont
  {O.}~\bibnamefont {Gutfleisch}}, \bibinfo {author} {\bibfnamefont
  {H.}~\bibnamefont {Kronm{\"{u}}ller}}, \bibinfo {author} {\bibfnamefont
  {G.}~\bibnamefont {Sch{\"{u}}tz}},\ and\ \bibinfo {author} {\bibfnamefont
  {E.~J.}\ \bibnamefont {Goering}},\ }\href
  {https://doi.org/10.1103/PhysRevB.95.024413} {\bibfield  {journal} {\bibinfo
  {journal} {Phys. Rev. B}\ }\textbf {\bibinfo {volume} {95}},\ \bibinfo
  {pages} {1} (\bibinfo {year} {2017})}\BibitemShut {NoStop}%
\bibitem [{\citenamefont {Pipich}\ and\ \citenamefont {Fu}(2015)}]{Pipich2015}%
  \BibitemOpen
  \bibfield  {author} {\bibinfo {author} {\bibfnamefont {V.}~\bibnamefont
  {Pipich}}\ and\ \bibinfo {author} {\bibfnamefont {Z.}~\bibnamefont {Fu}},\
  }\href {https://doi.org/10.17815/jlsrf-1-28} {\bibfield  {journal} {\bibinfo
  {journal} {Journal of Large-Scale Research Facilities}\ }\textbf {\bibinfo
  {volume} {1}},\ \bibinfo {pages} {A31} (\bibinfo {year} {2015})}\BibitemShut
  {NoStop}%
\bibitem [{\citenamefont {Pipich}(2018)}]{qtisas}%
  \BibitemOpen
  \bibfield  {author} {\bibinfo {author} {\bibfnamefont {V.}~\bibnamefont
  {Pipich}},\ }\href@noop {} {\bibinfo {title} {QtiSAS/QtiKWS Visualisation,
  Reduction, Analysis and Fit Framework with Focus on Small Angle
  Scattering}},\ \bibinfo {howpublished} {\url{http://qtisas.com}}
  (\bibinfo {year} {2018})\BibitemShut {NoStop}%
\bibitem [{\citenamefont {M{\"{u}}hlbauer}\ \emph {et~al.}(2016)\citenamefont
  {M{\"{u}}hlbauer}, \citenamefont {Heinemann}, \citenamefont {Wilhelm},
  \citenamefont {Karge}, \citenamefont {Ostermann}, \citenamefont {Defendi},
  \citenamefont {Schreyer}, \citenamefont {Petry},\ and\ \citenamefont
  {Gilles}}]{MUHLBAUER2016297}%
  \BibitemOpen
  \bibfield  {author} {\bibinfo {author} {\bibfnamefont {S.}~\bibnamefont
  {M{\"{u}}hlbauer}}, \bibinfo {author} {\bibfnamefont {A.}~\bibnamefont
  {Heinemann}}, \bibinfo {author} {\bibfnamefont {A.}~\bibnamefont {Wilhelm}},
  \bibinfo {author} {\bibfnamefont {L.}~\bibnamefont {Karge}}, \bibinfo
  {author} {\bibfnamefont {A.}~\bibnamefont {Ostermann}}, \bibinfo {author}
  {\bibfnamefont {I.}~\bibnamefont {Defendi}}, \bibinfo {author} {\bibfnamefont
  {A.}~\bibnamefont {Schreyer}}, \bibinfo {author} {\bibfnamefont
  {W.}~\bibnamefont {Petry}},\ and\ \bibinfo {author} {\bibfnamefont
  {R.}~\bibnamefont {Gilles}},\ }\href
  {https://doi.org/https://doi.org/10.1016/j.nima.2016.06.105} {\bibfield
  {journal} {\bibinfo  {journal} {Nucl. Instrum. Methods Phys. Res.~A}\
  }\textbf {\bibinfo {volume} {832}},\ \bibinfo {pages} {297 } (\bibinfo {year}
  {2016})}\BibitemShut {NoStop}%
\bibitem [{\citenamefont {Michels}(2014)}]{michels2014review}%
  \BibitemOpen
  \bibfield  {author} {\bibinfo {author} {\bibfnamefont {A.}~\bibnamefont
  {Michels}},\ }\href@noop {} {\bibfield  {journal} {\bibinfo  {journal} {J.
  Phys.: Condens. Matter}\ }\textbf {\bibinfo {volume} {26}},\ \bibinfo {pages}
  {383201} (\bibinfo {year} {2014})}\BibitemShut {NoStop}%
\bibitem [{\citenamefont {Michels}\ \emph {et~al.}(2014)\citenamefont
  {Michels}, \citenamefont {Erokhin}, \citenamefont {Berkov},\ and\
  \citenamefont {Gorn}}]{michels2014jmmm}%
  \BibitemOpen
  \bibfield  {author} {\bibinfo {author} {\bibfnamefont {A.}~\bibnamefont
  {Michels}}, \bibinfo {author} {\bibfnamefont {S.}~\bibnamefont {Erokhin}},
  \bibinfo {author} {\bibfnamefont {D.}~\bibnamefont {Berkov}},\ and\ \bibinfo
  {author} {\bibfnamefont {N.}~\bibnamefont {Gorn}},\ }\href@noop {} {\bibfield
   {journal} {\bibinfo  {journal} {J. Magn. Magn. Mater.}\ }\textbf {\bibinfo
  {volume} {350}},\ \bibinfo {pages} {55} (\bibinfo {year} {2014})}\BibitemShut
  {NoStop}%
\bibitem [{\citenamefont {Hammouda}(2010)}]{hammouda2010a}%
  \BibitemOpen
  \bibfield  {author} {\bibinfo {author} {\bibfnamefont {B.}~\bibnamefont
  {Hammouda}},\ }\href {https://doi.org/10.1107/S0021889810015773} {\bibfield
  {journal} {\bibinfo  {journal} {J. Appl. Crystallogr.}\ }\textbf {\bibinfo
  {volume} {43}},\ \bibinfo {pages} {716} (\bibinfo {year} {2010})}\BibitemShut
  {NoStop}%
\bibitem [{\citenamefont {Bender}\ \emph {et~al.}(2017)\citenamefont {Bender},
  \citenamefont {Bogart}, \citenamefont {Posth}, \citenamefont {Szczerba},
  \citenamefont {Rogers}, \citenamefont {Castro}, \citenamefont {Nilsson},
  \citenamefont {Zeng}, \citenamefont {Sugunan}, \citenamefont {Sommertune},
  \citenamefont {Fornara}, \citenamefont {Gonz{\'{a}}lez-Alonso}, \citenamefont
  {{Fern{\'{a}}ndez Barqu{\'{i}}n}},\ and\ \citenamefont
  {Johansson}}]{bender2017}%
  \BibitemOpen
  \bibfield  {author} {\bibinfo {author} {\bibfnamefont {P.}~\bibnamefont
  {Bender}}, \bibinfo {author} {\bibfnamefont {L.~K.}\ \bibnamefont {Bogart}},
  \bibinfo {author} {\bibfnamefont {O.}~\bibnamefont {Posth}}, \bibinfo
  {author} {\bibfnamefont {W.}~\bibnamefont {Szczerba}}, \bibinfo {author}
  {\bibfnamefont {S.~E.}\ \bibnamefont {Rogers}}, \bibinfo {author}
  {\bibfnamefont {A.}~\bibnamefont {Castro}}, \bibinfo {author} {\bibfnamefont
  {L.}~\bibnamefont {Nilsson}}, \bibinfo {author} {\bibfnamefont {L.~J.}\
  \bibnamefont {Zeng}}, \bibinfo {author} {\bibfnamefont {A.}~\bibnamefont
  {Sugunan}}, \bibinfo {author} {\bibfnamefont {J.}~\bibnamefont {Sommertune}},
  \bibinfo {author} {\bibfnamefont {A.}~\bibnamefont {Fornara}}, \bibinfo
  {author} {\bibfnamefont {D.}~\bibnamefont {Gonz{\'{a}}lez-Alonso}}, \bibinfo
  {author} {\bibfnamefont {L.}~\bibnamefont {{Fern{\'{a}}ndez
  Barqu{\'{i}}n}}},\ and\ \bibinfo {author} {\bibfnamefont {C.}~\bibnamefont
  {Johansson}},\ }\href@noop {} {\bibfield  {journal} {\bibinfo  {journal}
  {Sci. Rep.}\ }\textbf {\bibinfo {volume} {7}},\ \bibinfo {pages} {45990}
  (\bibinfo {year} {2017})}\BibitemShut {NoStop}%
\bibitem [{\citenamefont {Svergun}\ and\ \citenamefont
  {Koch}(2003)}]{svergun03}%
  \BibitemOpen
  \bibfield  {author} {\bibinfo {author} {\bibfnamefont {D.~I.}\ \bibnamefont
  {Svergun}}\ and\ \bibinfo {author} {\bibfnamefont {M.~H.~J.}\ \bibnamefont
  {Koch}},\ }\href@noop {} {\bibfield  {journal} {\bibinfo  {journal} {Rep.
  Prog. Phys.}\ }\textbf {\bibinfo {volume} {66}},\ \bibinfo {pages} {1735}
  (\bibinfo {year} {2003})}\BibitemShut {NoStop}%
\bibitem [{\citenamefont {Fritz}\ and\ \citenamefont
  {Glatter}(2006)}]{glatter2006}%
  \BibitemOpen
  \bibfield  {author} {\bibinfo {author} {\bibfnamefont {G.}~\bibnamefont
  {Fritz}}\ and\ \bibinfo {author} {\bibfnamefont {O.}~\bibnamefont
  {Glatter}},\ }\href@noop {} {\bibfield  {journal} {\bibinfo  {journal} {J.
  Phys.: Condens. Matter}\ }\textbf {\bibinfo {volume} {18}},\ \bibinfo {pages}
  {S2403} (\bibinfo {year} {2006})}\BibitemShut {NoStop}%
\bibitem [{\citenamefont {Lang}\ and\ \citenamefont
  {Glatter}(1996)}]{glatter1996}%
  \BibitemOpen
  \bibfield  {author} {\bibinfo {author} {\bibfnamefont {P.}~\bibnamefont
  {Lang}}\ and\ \bibinfo {author} {\bibfnamefont {O.}~\bibnamefont {Glatter}},\
  }\href@noop {} {\bibfield  {journal} {\bibinfo  {journal} {Langmuir}\
  }\textbf {\bibinfo {volume} {12}},\ \bibinfo {pages} {1193} (\bibinfo {year}
  {1996})}\BibitemShut {NoStop}%
\bibitem [{\citenamefont {Fritz-Popovski}\ \emph {et~al.}(2011)\citenamefont
  {Fritz-Popovski}, \citenamefont {Bergmann},\ and\ \citenamefont
  {Glatter}}]{glatter2011}%
  \BibitemOpen
  \bibfield  {author} {\bibinfo {author} {\bibfnamefont {G.}~\bibnamefont
  {Fritz-Popovski}}, \bibinfo {author} {\bibfnamefont {A.}~\bibnamefont
  {Bergmann}},\ and\ \bibinfo {author} {\bibfnamefont {O.}~\bibnamefont
  {Glatter}},\ }\href@noop {} {\bibfield  {journal} {\bibinfo  {journal} {Phys.
  Chem. Chem. Phys.}\ }\textbf {\bibinfo {volume} {13}},\ \bibinfo {pages}
  {5872} (\bibinfo {year} {2011})}\BibitemShut {NoStop}%
\bibitem [{\citenamefont {Kronm\"uller}\ and\ \citenamefont
  {F\"ahnle}(2003)}]{kronfahn03}%
  \BibitemOpen
  \bibfield  {author} {\bibinfo {author} {\bibfnamefont {H.}~\bibnamefont
  {Kronm\"uller}}\ and\ \bibinfo {author} {\bibfnamefont {M.}~\bibnamefont
  {F\"ahnle}},\ }\href@noop {} {\emph {\bibinfo {title} {Micromagnetism and the
  Microstructure of Ferromagnetic Solids}}}\ (\bibinfo  {publisher} {Cambridge
  University Press},\ \bibinfo {address} {Cambridge},\ \bibinfo {year}
  {2003})\BibitemShut {NoStop}%
\bibitem [{\citenamefont {Mettus}\ and\ \citenamefont
  {Michels}(2015)}]{Mettus2015}%
  \BibitemOpen
  \bibfield  {author} {\bibinfo {author} {\bibfnamefont {D.}~\bibnamefont
  {Mettus}}\ and\ \bibinfo {author} {\bibfnamefont {A.}~\bibnamefont
  {Michels}},\ }\href@noop {} {\bibfield  {journal} {\bibinfo  {journal} {J.
  Appl. Cryst.}\ }\textbf {\bibinfo {volume} {48}},\ \bibinfo {pages} {1437}
  (\bibinfo {year} {2015})}\BibitemShut {NoStop}%
\bibitem [{rgc()}]{rgcomment}%
  \BibitemOpen
  \href@noop {} {}\bibinfo {note} {Note that for these two samples the
  respective maximum of the $p(r)$ function, which is indicative of the
  ``particle'' radius $R$, roughly agrees with the $R_G$-value computed
  according to $R_G^2 = \frac{3}{5} R^2$ (assuming a spherical particle
  shape).}\BibitemShut {Stop}%
\bibitem [{\citenamefont {Curcio}\ \emph {et~al.}(2015)\citenamefont {Curcio},
  \citenamefont {Olivetti}, \citenamefont {Martino}, \citenamefont
  {K{\"{u}}pferling},\ and\ \citenamefont {Basso}}]{Curcio2015}%
  \BibitemOpen
  \bibfield  {author} {\bibinfo {author} {\bibfnamefont {C.}~\bibnamefont
  {Curcio}}, \bibinfo {author} {\bibfnamefont {E.~S.}\ \bibnamefont
  {Olivetti}}, \bibinfo {author} {\bibfnamefont {L.}~\bibnamefont {Martino}},
  \bibinfo {author} {\bibfnamefont {M.}~\bibnamefont {K{\"{u}}pferling}},\ and\
  \bibinfo {author} {\bibfnamefont {V.}~\bibnamefont {Basso}},\ }\href
  {https://doi.org/10.1016/j.phpro.2015.12.135} {\bibfield  {journal} {\bibinfo
   {journal} {Phys. Procedia}\ }\textbf {\bibinfo {volume} {75}},\ \bibinfo
  {pages} {1230} (\bibinfo {year} {2015})}\BibitemShut {NoStop}%
\bibitem [{\citenamefont {Zamora}\ \emph {et~al.}(2018)\citenamefont {Zamora},
  \citenamefont {Betancourt},\ and\ \citenamefont {Figueroa}}]{Zamora2018}%
  \BibitemOpen
  \bibfield  {author} {\bibinfo {author} {\bibfnamefont {J.}~\bibnamefont
  {Zamora}}, \bibinfo {author} {\bibfnamefont {I.}~\bibnamefont {Betancourt}},\
  and\ \bibinfo {author} {\bibfnamefont {I.~A.}\ \bibnamefont {Figueroa}},\
  }\href {https://doi.org/10.1007/s10948-017-4240-0} {\bibfield  {journal}
  {\bibinfo  {journal} {J. Supercond. Nov. Magn.}\ }\textbf {\bibinfo {volume}
  {31}},\ \bibinfo {pages} {873} (\bibinfo {year} {2018})}\BibitemShut
  {NoStop}%
\bibitem [{\citenamefont {Wroblewski}\ \emph {et~al.}(1999)\citenamefont
  {Wroblewski}, \citenamefont {Jansen}, \citenamefont {Sch\"afer},\ and\
  \citenamefont {Skowronek}}]{WROBLEWSKI1999}%
  \BibitemOpen
  \bibfield  {author} {\bibinfo {author} {\bibfnamefont {T.}~\bibnamefont
  {Wroblewski}}, \bibinfo {author} {\bibfnamefont {E.}~\bibnamefont {Jansen}},
  \bibinfo {author} {\bibfnamefont {W.}~\bibnamefont {Sch\"afer}},\ and\
  \bibinfo {author} {\bibfnamefont {R.}~\bibnamefont {Skowronek}},\ }\href@noop
  {} {\bibfield  {journal} {\bibinfo  {journal} {Nucl. Instrum. Methods Phys.
  Res.~A}\ }\textbf {\bibinfo {volume} {423}},\ \bibinfo {pages} {428}
  (\bibinfo {year} {1999})}\BibitemShut {NoStop}%
\bibitem [{\citenamefont {Yusuf}\ \emph {et~al.}(2006)\citenamefont {Yusuf},
  \citenamefont {De~Teresa}, \citenamefont {Mukadam}, \citenamefont
  {Kohlbrecher}, \citenamefont {Ibarra}, \citenamefont {Arbiol}, \citenamefont
  {Sharma},\ and\ \citenamefont {Kulshreshtha}}]{ibarra2006}%
  \BibitemOpen
  \bibfield  {author} {\bibinfo {author} {\bibfnamefont {S.~M.}\ \bibnamefont
  {Yusuf}}, \bibinfo {author} {\bibfnamefont {J.~M.}\ \bibnamefont
  {De~Teresa}}, \bibinfo {author} {\bibfnamefont {M.~D.}\ \bibnamefont
  {Mukadam}}, \bibinfo {author} {\bibfnamefont {J.}~\bibnamefont
  {Kohlbrecher}}, \bibinfo {author} {\bibfnamefont {M.~R.}\ \bibnamefont
  {Ibarra}}, \bibinfo {author} {\bibfnamefont {J.}~\bibnamefont {Arbiol}},
  \bibinfo {author} {\bibfnamefont {P.}~\bibnamefont {Sharma}},\ and\ \bibinfo
  {author} {\bibfnamefont {S.~K.}\ \bibnamefont {Kulshreshtha}},\ }\href@noop
  {} {\bibfield  {journal} {\bibinfo  {journal} {Phys. Rev. B}\ }\textbf
  {\bibinfo {volume} {74}},\ \bibinfo {pages} {224428} (\bibinfo {year}
  {2006})}\BibitemShut {NoStop}%
\end{thebibliography}
\bibliographystyle{apsrev4-2}

\end{document}